\documentclass[sigconf, acmsmall]{acmart}

\setcopyright{cc}
\setcctype{by}
\acmJournal{TODAES}
\acmPrice{}\acmDOI{10.1145/3769307}

\renewcommand{\shortauthors}{S. Pouget et al.}

\usepackage{algorithm2e}
\usepackage{listings}
\usepackage{xcolor}
\usepackage{dsfont}
\usepackage{geometry}
\usepackage{changepage}
\usepackage{forest}
\usepackage{tikz}
\usepackage{caption}
\usetikzlibrary{tikzmark}
\usepackage{tabularx}
\newcommand{\myparagraph}[1]{\par\textit{ #1}~}
\usepackage{multicol}
\usepackage{enumitem}
\usepackage{longtable}
\usepackage{supertabular}
\usepackage{xspace}
\usepackage{listings}
\usepackage{xcolor}
\usepackage{dsfont}

\usepackage{multirow}

\usepackage{tikz}

\usetikzlibrary{shapes.geometric, arrows.meta, positioning, backgrounds, fit, shadows.blur}
\usepackage{graphicx}

\definecolor{prometheusblue}{RGB}{39, 116, 174}
\definecolor{gold}{RGB}{255, 209, 0}

\usetikzlibrary{fit}

\usepackage{xcolor}

\newcommand{\newtext}[1]{{#1}}
\newcommand{\newtextt}[1]{{#1}}

\tikzset{
  process/.style = {rectangle, minimum width=3cm, minimum height=1.2cm, text centered, draw=black, fill=white,align=center},
  header/.style = {rectangle, text=gold, text centered, font=\bfseries\large, draw=none},
  arrow/.style = {thick,->,>=stealth},
  circleproc/.style = {circle, draw=black, fill=gray!50, minimum size=0.4cm, inner sep=0pt}
}

\usepackage{multicol}

\usepackage{xcolor}  
\usepackage{pifont}  

\newcommand{\yes}{\textcolor{green}{\ding{51}}}
\newcommand{\no}{\textcolor{red}{\ding{55}}}

\newcommand{\yesbad}{\textcolor{red}{\ding{51}}}
\newcommand{\nogood}{\textcolor{green}{\ding{55}}}

\definecolor{codegreen}{rgb}{0,0.6,0}
\definecolor{codegray}{rgb}{0.5,0.5,0.5}
\definecolor{codepurple}{rgb}{0.58,0,0.82}
\definecolor{backcolour}{rgb}{0.95,0.95,0.92}

\lstdefinestyle{cppstyle}{
    language=C++,
    commentstyle=\color{codegreen},
    keywordstyle=\color{blue},
    numberstyle=\tiny\color{codegray},
    stringstyle=\color{codepurple},
    basicstyle=\ttfamily\footnotesize,
    breakatwhitespace=false,
    breaklines=true,
    captionpos=b,
    keepspaces=true,
    numbers=left,
    numbersep=5pt,
    showspaces=false,
    showstringspaces=false,
    showtabs=false,
    tabsize=4
}

\newcommand{\framework}{Prometheus\xspace}
\newcommand{\git}{\href{https://github.com/UCLA-VAST/Prometheus}{\url{https://github.com/UCLA-VAST/Prometheus}}}
\usepackage[caption=false, font=normalsize, labelfont=sf, textfont=sf]{subfig}
\newcolumntype{C}[1]{>{\centering\arraybackslash}p{#1}}
\usepackage{adjustbox}

\begin{document}

\title{Holistic Optimization Framework for FPGA Accelerators}

\author{Stéphane Pouget}
\email{pouget@cs.ucla.edu}
\orcid{0000-0003-3950-5818}
\affiliation{%
  \institution{University of California, Los Angeles}
  \city{Los Angeles}
  \state{CA}
  \country{USA}
}

\author{Michael Lo}
\email{milo168@ucla.edu}
\orcid{0000-0002-8004-2942}
\affiliation{%
  \institution{University of California, Los Angeles}
  \city{Los Angeles}
  \state{CA}
  \country{USA}
}

\author{Louis-Noël Pouchet}
\email{pouchet@colostate.edu}
\orcid{0000-0001-5103-3097}
\affiliation{%
  \institution{Colorado State University}
  \city{Fort Collins}
  \state{CO}
  \country{USA}
}

\author{Jason Cong}
\email{cong@cs.ucla.edu}
\orcid{0000-0003-2887-6963}
\affiliation{%
  \institution{University of California, Los Angeles}
  \city{Los Angeles}
  \state{CA}
  \country{USA}
}

\renewcommand{\shortauthors}{Pouget et al.}

\begin{abstract}


Customized accelerators have revolutionized modern computing by delivering substantial gains in energy efficiency and performance through hardware specialization. Field-Programmable Gate Arrays (FPGAs) play a crucial role in this paradigm, offering unparalleled flexibility and high-performance potential. High-Level Synthesis (HLS) and source-to-source compilers have simplified FPGA development by translating high-level programming languages into hardware descriptions enriched with directives. However, achieving high Quality of Results (QoR) remains a significant challenge, requiring intricate code transformations, strategic directive placement, and optimized data communication.

This paper presents \textbf{Prometheus}, a holistic optimization framework that integrates key optimizations — including \textit{task fusion, tiling, loop permutation, computation-communication overlap, and concurrent task execution}—into a unified design space. By leveraging \textit{Non-Linear Programming (NLP) methodologies}, Prometheus explores the optimization space under strict resource constraints, enabling automatic bitstream generation. Unlike existing frameworks, Prometheus considers interdependent transformations and dynamically balances computation and memory access.

We evaluate Prometheus across multiple benchmarks, demonstrating its ability to maximize parallelism, minimize execution stalls, and optimize data movement. The results showcase its superior performance compared to state-of-the-art FPGA optimization frameworks, highlighting its effectiveness in delivering high QoR while reducing manual tuning efforts.

\end{abstract}


\received{20 February 2007}
\received[revised]{12 March 2009}
\received[accepted]{5 June 2009}

\maketitle

\section{Introduction}
\label{introduction}

The rise of customized accelerators has revolutionized modern computing, driving significant improvements in energy efficiency and performance through hardware specialization. Among these accelerators, {Field-Programmable Gate Arrays (FPGAs)} have gained widespread adoption due to their {flexibility, high performance, and adaptability} across diverse domains, including machine learning, scientific computing, financial modeling, and embedded systems. Unlike fixed-function hardware such as {Application-Specific Integrated Circuits (ASICs)}, FPGAs offer {reconfigurability}, enabling hardware adaptation to workload-specific requirements. However, achieving a good \textit{Quality of Result} (QoR) on FPGAs remains a {complex and resource-intensive process}, requiring careful tuning of {computation, memory access, and parallelism strategies}.

To address this complexity, {High-Level Synthesis (HLS) tools \cite{ vitis, intel_fpga, catapult, legup} and source-to-source compilers} e.g, ~\cite{ scalehls, heterocl, heterohalide, sisyphus, stream_hls}, have emerged as {critical enablers of FPGA adoption}, allowing developers to write {hardware-accelerated programs} in high-level languages such as C++ and Python. These tools generate synthesizable hardware descriptions, leveraging {pragmas (hardware directives) and code transformations} to optimize execution. Despite their advantages, {HLS-generated designs still require substantial manual tuning} to achieve high QoR, as the compiler’s ability to optimize performance remains highly dependent on the {structure and directives applied to the input code}.

\subsection{Challenges in FPGA Optimization}

Despite significant advancements in {HLS automation} and {FPGA design methodologies}, achieving high-performance and resource-efficient FPGA implementations remains a challenging problem. This complexity arises from the intricate interactions between computation, memory access, and parallel execution strategies. Several key challenges must be addressed to unlock the full potential of FPGA acceleration.

\textbf{Code Transformation and Design Space Exploration:}  
FPGA performance is highly dependent on {how computations are structured and scheduled}. Optimizations such as {loop tiling, loop permutation, and loop unrolling} play a crucial role in exposing parallelism and improving memory access efficiency. 

\textbf{Task Concurrency and Resource Management:}  
Maximizing throughput in FPGA designs requires efficient {task concurrency}, where multiple computation kernels execute in parallel. However, coordinating parallel tasks introduces challenges in managing {resource contention, memory bandwidth, and routing complexity}. Overlapping computations effectively demands intelligent scheduling strategies that minimize {stall cycles} while avoiding bottlenecks caused by {limited on-chip storage} and {off-chip memory latency}. 

\textbf{Computation-Communication Overlap:}  
FPGA acceleration is often hindered by inefficient data movement between {off-chip memory (DDR/HBM)} and {on-chip compute units}. Many workloads experience severe performance degradation due to memory access bottlenecks rather than computational limitations. {Dataflow architectures} aim to address this by implementing {FIFO-based streaming} to pipeline data transfers, but they often fail to exploit {intra-task parallelism} efficiently. In contrast, {shared buffer models}~\cite{sisyphus, scalehls} improve {data reuse}, but their rigid memory management strategies make it difficult to {overlap computation and communication dynamically}. A more adaptive approach is needed to {orchestrate data movement in parallel with execution}, reducing overall memory stalls.

\textbf{Scalability and Multi-\newtext{Super Logic Regions Utilization:}}  
FPGA architectures, especially those using {Stacked Silicon Interconnect (SSI) technology}, feature multiple {Super Logic Regions (SLRs)} to enhance scalability. However, most frameworks, including {Merlin-based tools} 
{restrict designs to a single SLR}, 
leading to {underutilization of FPGA resources}. While designing across multiple SLRs is possible, it significantly increases {routing congestion, timing closure failures, and bitstream generation complexity}. Without an SLR-aware optimization strategy, large-scale tasks face a high probability of {bitstream generation failures} or degraded performance due to excessive {inter-SLR communication overhead}.

\textbf{Comprehensive Design Space Exploration:}  
Current design space exploration (DSE) methodologies are often {narrow in scope}, considering only a limited subset of optimizations at a time. Each transformation—whether it involves loop reordering, tiling, pipelining, or data partitioning—has a profound impact on the overall performance and resource usage. 
Frameworks that optimize only one aspect of the design fail to account for the interdependencies between different optimizations, leading to suboptimal results. To achieve maximum efficiency, FPGA designs require a {global optimization approach} that integrates {all possible transformations} while considering the constraints imposed by memory hierarchy, computation parallelism, and hardware utilization.

Addressing these challenges requires a unified optimization strategy that seamlessly integrates code transformation, task concurrency management, computation-communication overlap, and hardware-aware scheduling into a single framework. The proposed work aims to develop such an approach, ensuring high-performance FPGA implementations while maintaining synthesis feasibility.

\subsection{Contributions and \framework{} Method}

This work introduces \framework{}, a unified framework that automates FPGA design space exploration by integrating multiple optimizations into a single methodology 
\newtext{
specialized 
\emph{for affine programs} \cite{mlir}. 
Specifically, we operate on affine loop nests where statements can be distributed in multiple loops (i.e. loops are permutable \cite{pouchet.11.popl}).
Such programs encompass typical computational pattern in dense linear algebra, data mining, image processing, etc. as exemplified in the PolyBench benchmarking suite \cite{polybench, polybench-python}.}
\newtextt{Unlike existing tools that optimize isolated aspects of FPGA acceleration such as {AutoDSE \cite{autodse}, HARP \cite{harp}, NLP-DSE \cite{nlp_dse}}, {Sisyphus} \cite{sisyphus} and Stream-HLS \cite{stream_hls}}, \framework{} holistically addresses {code transformation, memory management, task concurrency, and hardware-aware scheduling}. It leverages a {hybrid execution model} that balances {shared memory reuse} and {dataflow streaming} to maximize parallelism while minimizing memory overhead. 

To explore the vast design space efficiently, \framework{} formulates the optimization process as a {Non-Linear Programming (NLP)} problem, enabling automatic selection of {loop transformations, tiling strategies, pragma configurations, and memory partitioning policies}. Unlike heuristic-based approaches, this formulation captures complex {optimization interdependencies} and ensures globally efficient designs. Additionally, \framework{} incorporates {SLR-aware task scheduling}, which partitions tasks across {multiple Super Logic Regions (SLRs)}. This addresses a major limitation in prior works that {restrict execution to a single SLR}, leading to underutilization of FPGA resources. By integrating SLR-aware optimizations, \framework{} reduces {routing congestion}, improves {scalability}, and ensures {successful bitstream generation}.

The methodology follows a structured process: First, \framework{} performs \textit{affine code analysis} to identify parallelism opportunities and dependencies. It then generates a \textit{dataflow graph} and fuses tasks that share the same outputs to minimize unnecessary communication overhead. 
Next, \framework{} constructs a comprehensive design space that includes all key optimizations outlined in Section~\ref{sec:motivation}, such as \textit{task concurrency via dataflow}, \textit{computation-communication overlap}, and \textit{adaptive parallelism}, while ensuring resource constraints are met for each SLR. 
The \textit{NLP-based design space exploration model}, as described in Section~\ref{sec:nlp}, is then employed to identify the theoretically optimal set of transformations. Once the best configuration is identified, \framework automatically generates the optimized design. Finally, the system produces \textit{OpenCL host code} and compiles the FPGA bitstream with the HLS compiler.

The key contributions of this work are:

\begin{itemize}
    \item {Unified FPGA optimization framework} that jointly optimizes {loop transformations, pragma insertion, task concurrency and computation-communication overlap} while considering interdependencies between computation and data movement.
    \item {Hybrid execution model} that dynamically selects between {shared buffering} and {dataflow streaming} to maximize {parallelism} and {efficient memory access}.
    \item {NLP-based design space exploration} that automates the selection of {loop tiling, scheduling, and memory strategies} to achieve a globally optimal theoretical performance.
    \item {SLR-aware scheduling and multi-SLR partitioning} to balance {routing complexity} and {FPGA resource utilization}, overcoming single-SLR limitations in prior works.
    \item \newtext{End-to-end compilation and automation, generating optimized \textit{HLS-C++ code}—that is, standard C++ source code annotated with AMD/Vitis-specific high-level synthesis (HLS) pragmas to guide hardware generation—alongside OpenCL host code and FPGA bitstreams, all with minimal manual intervention.}
    \item {Comprehensive performance evaluation}, demonstrating superior QoR compared to {AutoDSE}~\cite{autodse}, {Sisyphus}~\cite{sisyphus}, {ScaleHLS}~\cite{scalehls}, Stream-HLS \cite{stream_hls} and Allo \cite{allo}.
\end{itemize}

\newtext{Our framework, Prometheus, is open source and available at \git.}



\section{Background and Motivation}

\label{sec:motivation}

Optimizing latency and performance in FPGA designs relies on a combination of code transformations and strategic insertion of directives within High-Level Synthesis (HLS) \cite{DBLP:journals/corr/abs-1807-01340}. This section explores various HLS optimization techniques, their role in enhancing parallelism and resource utilization, as well as their limitations and challenges.

Table \ref{tab:related} presents a comparative analysis of various frameworks built on top of AMD Vitis HLS. These frameworks are categorized based on their underlying methodologies, including Model-Free, AI-Based, Cost Model Communication, Cost Model Computation, and NLP-Based approaches. The table evaluates their capabilities across key optimization techniques and objectives.

\begin{table}[!htb]
\footnotesize{
    \centering
    \begin{tabular}{
    @{}p{0.15\linewidth} 
    p{0.06\linewidth}
    p{0.06\linewidth}
    p{0.06\linewidth}
    p{0.06\linewidth}
    p{0.06\linewidth}
    p{0.06\linewidth}
    p{0.06\linewidth}
    p{0.08\linewidth}
    p{0.10\linewidth} 
    @{}
    }
        \toprule
        & Model Free & AI Model & Cost Model Communication & Cost Model Computation & \multicolumn{4}{c}{NLP-Based} \\
        \cmidrule(lr){2-2}
        \cmidrule(lr){3-3}
        \cmidrule(lr){4-4}
        \cmidrule(lr){5-5}
        \cmidrule(lr){6-6} 
        \cmidrule(lr){7-10} 
        & AutoDSE \cite{autodse} & HARP \cite{harp} & PolyOpt-HLS \cite{pouchet:fpga13} & ScaleHLS \cite{scalehls} / POM~\cite{pom} & HeteroCL \cite{heterocl} / Allo \cite{allo} & Stream HLS \cite{stream_hls} & NLP-DSE \cite{nlp_dse} & Sisyphus \cite{sisyphus} & Prometheus \\
        \midrule
        Pragma Insertion & \yes & \yes & Limit & \yes & \yes & \yes & \yes & \yes & \yes \\
        \cmidrule(lr){0-0}
        Code Transformation (Tiling) & \no & \no & \yes & Limit & \yes & Limit & \no & \yes & \yes \\
        \cmidrule(lr){0-0}
        Code Transformation (Loop Permutation) & \no & \no & Limit & Limit & \yes & \yes & \no & \yes & \yes \\

        \cmidrule(lr){0-0}
        Code transformation + pragma insertion (unified) & \no & \no & \yes & \no & \no & \no & \no & \yes & \yes \\
        \cmidrule(lr){0-0}
        Task Concurrency & \no & \no & \yes & Limit & \no & \yes & \no & \no & \yes \\
        \cmidrule(lr){0-0}
        Dataflow & \no & \no & \yes & \no & \yes & \yes & \no & \no & \yes \\
        \cmidrule(lr){0-0}

        Computation-Communication Overlap & \no & \no & \yes & Limit & \no  & \no & \no & \no & \yes \\
        \cmidrule(lr){0-0}

        Data Packing & \yes & \yes & \yes & \no & \yes & \no & \yes & \yes & \yes \\
        \cmidrule(lr){0-0}

        Padding (Communication) & \no & \no & \yes & \no & \no & \no & \no & \no & \yes \\
        \cmidrule(lr){0-0}

        Padding (Computation) & \no & \no & \no & \no & \no & \no & \no & \no & \yes \\
        \cmidrule(lr){0-0}

        Hardware Aware & \no & \no & \no & \no & \no & \no & \no & Limit & \yes \\
        \cmidrule(lr){0-0}
        Hardware-feasible & Limit & Limit & \no & \no & \no & \no & Limit & Limit & \yes \\
        \cmidrule(lr){0-0}
        
        Management of Off-Chip and On-Chip Memory Transfers & \yes & \yes & \yes & \no & \yes & \no & \yes & \yes & \yes \\
        \cmidrule(lr){0-0}

        Objective & Comm + Comp & Comm + Comp & Comm & Comp & - & Comp & Comm + Comp & Comm + Comp & Comm + Comp \\
        \cmidrule(lr){0-0}
        
        Enumeration (AI, heuristics, …) & \yesbad & \yesbad & \nogood   & \yesbad & Manual & \nogood & \nogood & \nogood & \nogood  \\

        \bottomrule
    \end{tabular}
    }
\caption{\textit{Comparison of HLS-Based FPGA Optimization Frameworks.}
This table summarizes key features supported by various frameworks, including pragma insertion, code transformations, task concurrency, dataflow modeling, and hardware-awareness, categorized by their underlying optimization strategies.}
    \label{tab:related}
\end{table}

\subsection{HLS Optimization}

\subsubsection{Pragma Insertion}

HLS optimizations rely extensively on a range of directives that control loop execution, data access patterns, and computation scheduling to maximize performance. Among these, three fundamental pragmas—\textit{unroll}, \textit{pipeline}, and \textit{array partitioning}—are particularly critical for improving parallelism and reducing execution latency.

The \textit{unroll} directive enables the expansion of loop iterations in the hardware design, allowing multiple iterations to execute in parallel. By eliminating loop control overhead and enabling greater resource utilization, unrolling can significantly boost throughput. However, its effectiveness is limited by FPGA resource constraints, such as available DSP slices, BRAM, and routing congestion.

The \textit{pipeline} pragma restructures loop execution to allow overlapping operations, thereby reducing the initiation interval (II)—the number of cycles required to launch consecutive iterations. This directive is essential for achieving high throughput in applications with iterative computations, such as matrix multiplications and stencil operations. Careful tuning of pipeline depth and initiation interval is necessary to balance resource usage while avoiding performance bottlenecks due to memory access contention.

\textit{Array partitioning} is another crucial optimization that enhances memory access parallelism. By splitting large arrays into smaller banks stored in separate on-chip BRAM blocks or distributed memory structures, this pragma allows simultaneous data accesses, facilitating more effective unrolling and pipelining. Without partitioning, memory conflicts and bandwidth limitations can restrict performance, especially with multiple concurrent memory accesses.

Beyond these fundamental pragmas, additional directives such as \textit{loop flattening}, and \textit{dependency pragmas} can further optimize execution. \textit{Loop flattening} combines nested loops into a single loop to improve hardware efficiency. Dependency pragmas help manage data dependencies to ensure efficient scheduling without unnecessary synchronization delays.

Pragma insertion has been extensively studied using various methodologies to optimize performance and resource utilization in HLS designs. These approaches include bottleneck analysis, as employed by AutoDSE \cite{autodse}, performance estimation through Graph Neural Networks (GNN) in frameworks like HARP \cite{harp}, and cost model optimization leveraging Non-Linear Programming (NLP) in NLP-DSE \cite{nlp_dse, nlp_dse_poster}.

Each method presents distinct advantages and limitations. AutoDSE achieves full accuracy by running the HLS compiler for each configuration, ensuring precise performance measurements. However, this approach is computationally expensive and time-consuming, making it impractical for rapid design exploration. HARP, on the other hand, can estimate performance across configurations without direct compilation, significantly reducing evaluation time. However, it requires extensive training datasets and fine-tuning for each new kernel to accurately model compiler behavior. While this approach enables learning-based adaptation to different designs, its effectiveness depends heavily on data quality and model calibration.
Conversely, NLP-DSE explores the entire design space efficiently by formulating the optimization as a Non-Linear Programming problem. This method allows for rapid exploration and selection of theoretical optimal configurations. However, it relies on a theoretical cost model, which may introduce inaccuracies if the compiler behaves unpredictably or if certain optimizations are not accurately captured by the model. Consequently, while NLP-DSE provides a fast and scalable solution, its reliability depends on the fidelity of the cost model to actual HLS compilation behavior.

While pragma insertion is a powerful optimization technique, its effectiveness is inherently dependent on the interaction between memory hierarchy, computational resources, and FPGA-specific constraints. Moreover, its impact is tightly coupled with the program’s execution schedule, meaning that without appropriate code transformations, the benefits of pragma insertion remain limited. Effective optimization requires a synergy between pragma directives and structural modifications to the code to fully exploit FPGA parallelism and resource utilization.

\subsubsection{Code Transformation}

Code transformations are fundamental for optimizing execution on FPGAs. While pragma insertion plays a crucial role in enhancing performance, it must be complemented by effective code transformations to fully exploit FPGA resources. Techniques such as loop reordering (permutation), task fusion, and data tiling enhance data locality and increase parallelism, thereby improving both computational efficiency and memory access patterns.

Various code transformation strategies have been developed specifically for FPGAs \cite{pouchet.fpga.13, zhaopolsca,zhao2021phism, li2014throughput, liu2016loop, liu2017polyhedral, 7160061, choi.iccad.18}. 
While transformations originally designed for CPUs and GPUs achieve substantial performance gains by optimizing for their respective architectures, they do not inherently align with FPGA requirements, which prioritize fine-grained parallelism and efficient resource utilization.
Several studies have utilized Pluto \cite{pluto}, a leading compiler framework originally designed for CPU optimizations, to transform FPGA kernels \cite{zhaopolsca, zhao2021phism}. While Pluto excels at tiling and minimizing dependencies to enhance memory reuse, its direct application to FPGA optimization is constrained due to the fundamental differences in optimization strategies required for CPUs and FPGAs. Unlike CPUs, where memory hierarchy and cache locality are primary concerns, FPGA optimization focuses on maximizing parallelism, minimizing resource contention, and efficiently utilizing on-chip memory, making Pluto's conventional transformation techniques less effective in this context.
Conversely, studies such as \cite{pouchet.fpga.13, li2014throughput, liu2016loop, liu2017polyhedral, 7160061, choi.iccad.18} focus on code transformations tailored to specific FPGA performance goals. The work in \cite{pouchet.fpga.13} aims to reduce communication overhead between off-chip and on-chip memory, achieving superior Quality of Results (QoR) for memory-bound kernels. Meanwhile, research efforts in \cite{li2014throughput, liu2016loop, liu2017polyhedral, 7160061, choi.iccad.18} concentrate on optimizing pipelining strategies to maximize instruction-level parallelism and resource utilization.
Sisyphus \cite{sisyphus} introduces a unified approach by integrating code transformation and pragma insertion into a single optimization problem. By formulating this as a Non-Linear Programming (NLP) problem, Sisyphus efficiently explores the design space to identify theoretical optimal configurations, streamlining FPGA acceleration while maintaining a balance between computation and memory access efficiency.

\subsubsection{Task Concurrency}

HLS tools like Vitis HLS support the \textit{dataflow} pragma, which structures computations into 
actors that communicate through FIFO queues. This approach allows overlapping execution of multiple tasks, significantly reducing overall latency 
. 
By enforcing a producer-consumer model, dataflow scheduling enables each task to process data as soon as it becomes available, rather than waiting for an entire stage to complete, which is particularly beneficial.

For computational kernels such as \textit{3mm} (Listing \ref{lst:3mm_motiv}), the dataflow paradigm allows the first two matrix multiplications to execute in parallel while the third begins as soon as its required inputs are produced. This overlapping of execution helps maximize throughput and minimize idle time for computing units. Additionally, dataflow optimizations facilitate task-level parallelism, allowing independent tasks to run concurrently across multiple compute resources, such as DSP blocks and BRAM, ensuring efficient utilization of available FPGA resources.

However, despite its advantages, pure dataflow parallelism presents several challenges. One major limitation is intra-task parallelism, which is constrained by the reliance on FIFO-based communication. Since each FIFO can only transfer up to 512 bits per cycle  (as this is the maximum off-chip memory bitwidth supported), this inherently restricts the amount of data that can be processed concurrently within a task. To further enhance intra-task parallelism, alternative approaches must be explored.

Stream-HLS \cite{stream_hls} addressed this limitation by increasing the number of FIFOs connecting two tasks, assuming that all data resides on-chip. While this approach simplifies certain aspects of the design space, it is neither scalable nor generalizable for transferring data from off-chip to on-chip memory. By using $n$ FIFOs to transfer $n$ data elements in parallel between two tasks, the method significantly increases resource consumption without proportionally improving efficiency. The additional FIFOs complicate the design, potentially leading to routing congestion and excessive hardware overhead.

Moreover, modern FPGAs feature up to 32 off-chip memory banks, making the multi-FIFO approach impractical for handling off-chip data transfers. A more effective and scalable strategy must be developed to optimize intra-task parallelism while ensuring efficient communication between off-chip and on-chip memory, without introducing unnecessary complexity or resource constraints.

\subsubsection{Shared Buffering}

Shared buffering is a critical technique in FPGA memory optimization, enabling efficient data access and reuse across multiple computational units. It can be employed at a global level, as seen in frameworks like AutoDSE \cite{autodse}, Sisyphus \cite{sisyphus}, and ScaleHLS \cite{scalehls}, or within individual dataflow tasks to enhance execution efficiency.

This approach involves preloading data buffers into on-chip memory, allowing multiple computations to access shared data without redundant transfers. By reducing memory access latency and improving bandwidth utilization, shared buffering facilitates high parallelism and optimizes overall performance. It is particularly beneficial in applications that are computation-bound.

However, shared buffering presents challenges in maintaining concurrency. While it enables efficient data reuse, it can introduce synchronization overhead, especially when multiple tasks attempt to access the same memory region simultaneously. Managing concurrent access requires arbitration mechanisms, which can lead to increased latency and potential bottlenecks. Additionally, routing congestion can occur due to high interconnect demands, limiting scalability in complex FPGA designs.

One limitation of shared buffering is its impact on initiation interval (II) in pipelined architectures. Unlike dataflow-based designs that rely on FIFO queues for seamless data streaming, shared buffering requires explicit read and write coordination, which may introduce stalls if not carefully managed. Moreover, FPGA resource constraints, such as limited BRAM and URAM availability, impose restrictions on buffer size and allocation strategies.

To address these challenges, advanced techniques such as double-buffering, memory partitioning, and adaptive scheduling have been explored. Double-buffering enables overlapping computation with data transfer, reducing idle cycles. Memory partitioning distributes data across multiple banks to alleviate contention, while adaptive scheduling dynamically assigns buffer access based on task priority and workload demands.

By effectively integrating shared buffering with intelligent memory management strategies, FPGA designs can achieve a balance between computational parallelism and efficient memory access. Future advancements should focus on automated buffer allocation techniques and dynamic access pattern optimization to further enhance performance in high-level synthesis (HLS) workflows.

\subsubsection{Computation-Communication Overlap}

Overlapping computation and communication is crucial for high-performance FPGA designs. Techniques such as double buffering (ping-pong buffering) and advanced data tiling help mask communication latency while keeping computational units busy. Managing data transfers efficiently between on-chip and off-chip memory ensures that processing units remain active without waiting for data.

\subsubsection{Data Packing and Padding}

Modern FPGA architectures support high-bandwidth data transfers (up to 512-bit wide on AMD/Xilinx FPGAs). Data packing and padding optimize memory alignment, reducing the number of required memory cycles. For instance, transferring 216 floating-point values using a 256-bit width (8 floats per transfer) requires 27 cycles -- compared to 216 cycles without packing. This highlights the importance of efficient data packing to minimize overhead. However, to enable such packing, the transfer vector size must evenly divide the total array size.

To fully exploit data packing, padding must be considered to enable even faster data transfers. Additionally, padding is valuable for achieving finer control over parallelism and resource utilization by expanding the available design space for loop unrolling, thereby enhancing computational throughput and efficiency.




\paragraph{Padding for Communication}

Padding must be considered to increase the flexibility of data packing. The original size of the data may impose restrictions on packing efficiency, but by introducing padding, we create a larger space that may allow for a more efficient transfer. However, padding is not a free optimization, as increasing it also increases the amount of data that needs to be transferred.


\begin{figure}[!htb]
    \centering
    \begin{minipage}{0.45\linewidth}
        \centering
        \includegraphics[width=\linewidth]{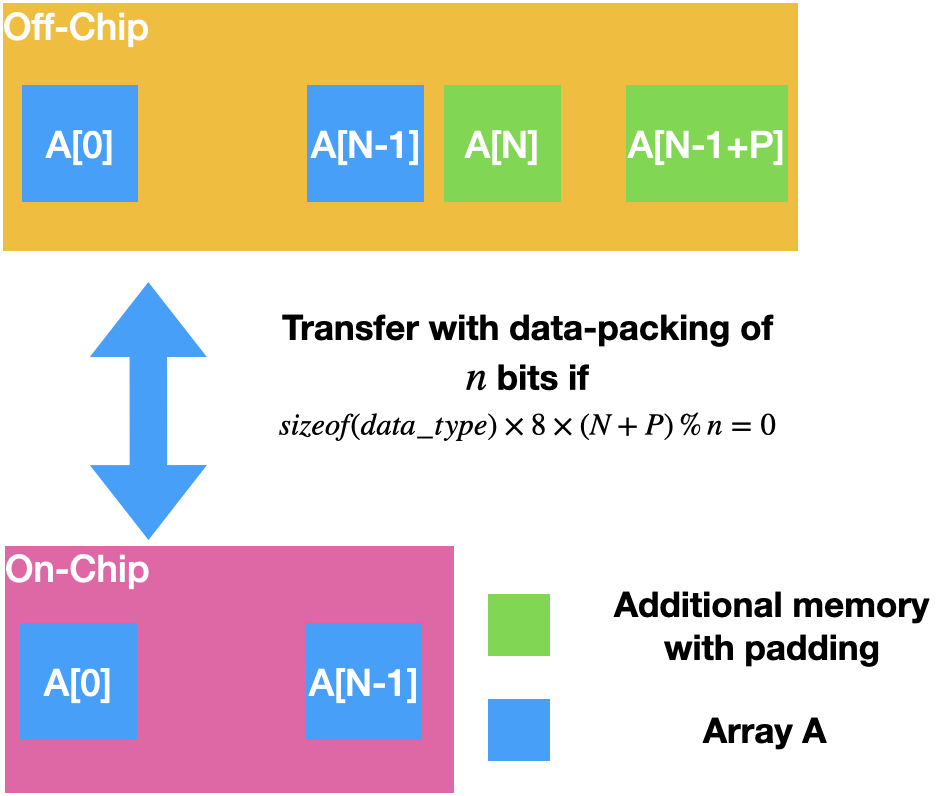}
        \caption{Illustration of Padding Strategy to Align Array Size for Efficient Data Packing and Memory Transfers on AMD FPGAs}
        \label{fig:comm-pad}
    \end{minipage}%
    \hfill
    \begin{minipage}{0.49\linewidth}
        \centering
        \begin{lstlisting}[language=C++, caption={Effect of Padding on the Space of Unroll Factors for Computation}, label={lst:padding}]
// Original loop without padding
for (int i = 0; i < 190; i++){
#pragma HLS unroll factor=UF
// Possible UF values: 1, 2, 5, 10, 19, 38, 95, 190
    A[i] += 1;
}      

// Padded loop 
for (int i = 0; i < 190 + 2; i++){ // Padding of 2
#pragma HLS unroll factor=UF
// Possible UF values: 1, 2, 3, 4, 6, 8, 12, 16, 24, 32, 48, 64, 96, 192
    A[i] += 1;
}      
        \end{lstlisting}
    \end{minipage}
\end{figure}

Figure \ref{fig:comm-pad} illustrates the selection of the padding value $P$ for an array $A$ of size $N$. By leveraging data packing (up to 512 bits), multiple elements can be transferred per cycle. However, due to constraints in AMD compilers and FPGA architectures, the total transfer size must be divisible by a power of two. Padding is therefore introduced to align the data size accordingly and enable more efficient transfers.

In the code presented in Listings \ref{lst:mm_orig} and \ref{lst:mm_tiled}, where $J = 190$, transferring all elements along the second dimension (iterated by loop $j$) of array $B$ onto the chip is constrained by the available memory bandwidth. Without padding, the maximum transfer rate is $64$ bits per cycle, as $190 \times 32$ is divisible by $64$ but not by $128$.
However, by introducing padding with $P = 2$ and adjusting $J$ to $192$ ($190 + 2$), the data alignment enables a significantly higher transfer rate of $512$ bits per cycle, as $192 \times 32$ is now perfectly divisible by $512$. This optimization maximizes memory bandwidth utilization, reducing transfer latency and improving overall data throughput.

\paragraph{Padding for Computation}

Padding can also be used for computation, if we take a similar method than Sisyphus \cite{sisyphus} which tiled and unroll the intra-tile loops which correspond to the transformation from Listing \ref{lst:mm_orig} to Listing \ref{lst:mm_tiled}. We will have an unroll factor which correspond to $I1 \times J1 \times K1$ but we want to have $I1$ divide $I$, $J1$ divide $J$ and $K1$ divide K because we do not want to use extra resource to only compute a partial tile which would correspond to the execution of the last tile which is not complete. For example if we have a trip count of $190$ and we use an unroll factor of $8$ then we will execute $184 \times 8$ a full tile and then 6 iterations for the partial tile.


If the unroll factor is restricted to be a divisor of the loop trip count, the design space becomes significantly limited. Padding can be used to adjust the trip count, thereby expanding the set of valid unroll factors to better match available hardware resources.
As shown in Listing~\ref{lst:padding}, a loop with a trip count of 190 permits only a limited set of unroll factors:
$
UF \in \{1, 2, 5, 10, 19, 38, 95, 190\}.
$
However, by padding the loop to a trip count of 192, the space of legal unroll factors becomes:
$
UF \in \{1, 2, 3, 4, 6, 8, 12, 16, 24, 32, 48, 64, 96, 192\}.$
This increased flexibility allows finer control over unrolling and resource utilization, enabling more efficient hardware implementations and better exploitation of computational parallelism.

\begin{figure}[!htb]
\begin{lstlisting}[label={lst:mm_orig},caption={\textit{Baseline Implementation of Matrix Multiplication in C.} This code depicts a naive triple-nested loop structure used for matrix-matrix multiplication, serving as the unoptimized reference for further transformations.},escapeinside={(*@}{@*)}]
for (i = 0; i < I; i++)  
  for (j = 0; j < J; j++) 
    for (k = 0; k < K; k++) 
      C[i][j] += A[i][k] * B[k][j];
\end{lstlisting}
\end{figure}

\begin{figure}[!htb]
\begin{lstlisting}[label={lst:mm_tiled},caption={\textit{Matrix Multiplication with Loop Tiling and Fully Unrolled Intra-Tile Computation.}
This implementation showcases a performance-optimized matrix multiplication using loop tiling to enhance data locality and fully unrolled intra-tile loops to expose fine-grained parallelism. },escapeinside={(*@}{@*)}]
for (i0 = 0; i0 < I0; i0++)  
  for (j0 = 0; j0 < J0; j0++) 
    for (k0 = 0; k0 < K0; k0++) 
      for (i1 = 0; i1 < I1; i1++)  
#pragma HLS unroll
        for (j1 = 0; j1 < J1; j1++) 
#pragma HLS unroll
          for (k1 = 0; k1 < K1; k1++) 
#pragma HLS unroll
            C[i0*I1+i1][j0*J1+j1] += A[i0*I1+i1][k0*K1+k1]*B[k0*K1+k1][j0*J1+j1];
\end{lstlisting}
\end{figure}

\subsection{Performance and Hardware Considerations}

\subsubsection{Performance Evaluation}

Performance evaluation in HLS is a multi-stage process that assesses different aspects of a design's efficiency and feasibility. It is typically conducted at three levels: estimation (e.g., Vitis HLS reports), RTL simulation, and FPGA-based evaluation. Each of these stages provides critical insights, but they vary in accuracy, speed, and resource requirements.

\begin{enumerate}

    \item  Estimation (HLS Reports): The first level of evaluation relies on estimation tools provided by HLS compilers such as Vitis HLS. These reports generate quick approximations of key performance metrics, including latency, resource utilization (DSPs, LUTs, BRAMs), and throughput. While these estimations are useful for early-stage design exploration, they do not account for low-level placement and routing effects, which can significantly impact real-world performance. The optimistic assumptions of HLS reports often fail to capture routing congestion, memory access delays, and other architectural bottlenecks.

    \item  RTL Simulation: To achieve more accurate performance insights, RTL simulation is performed after HLS compilation. This stage involves generating synthesizable RTL code, which is then simulated using hardware description language (HDL) tools. RTL simulation provides cycle-accurate performance metrics, allowing designers to analyze pipeline behavior, data dependencies, and execution timing. However, while RTL simulation models hardware more precisely than HLS reports, it does not incorporate real-world FPGA constraints such as clock tree synthesis, interconnect delays, and power distribution, which can affect the final implementation.

    \item  FPGA-Based Evaluation: The most accurate method of performance evaluation involves deploying the design onto an actual FPGA. This process includes place-and-route, bitstream generation, and execution on hardware. FPGA-based evaluation provides real execution metrics such as operating frequency, power consumption, and effective memory bandwidth. It also reveals practical challenges, including timing closure issues, placement inefficiencies, and resource contention that cannot be detected in earlier stages. This level of evaluation is essential for validating Quality of Results (QoR) and ensuring the design meets real-world constraints.
\end{enumerate}

The transition from RTL simulation to FPGA implementation often introduces additional challenges, such as increased routing complexity, unexpected timing violations, and suboptimal resource utilization. These discrepancies arise due to the abstraction gap between HLS-generated code and final FPGA hardware constraints. To mitigate these issues, it is essential to develop robust tools that enable rapid design regeneration by modifying specific configurations within a well-defined portion of the design space. 
Such tools would enable engineers to efficiently iterate on design parameters without requiring a complete code regeneration, which can be time-consuming and may result in a significantly different design that still fails to generate a valid bitstream. Having the ability to selectively modify only the congested parts of the design while preserving the rest of the configuration would be highly valuable, ensuring faster convergence toward an optimized and feasible FPGA implementation.

\subsubsection{Hardware Awareness and Resource Constraints}

Many existing studies conclude their evaluations at the Vitis HLS report or RTL simulation stage, overlooking the critical impact of placement and routing constraints. However, as the design progresses toward bitstream generation, the available design space becomes increasingly constrained. Addressing hardware limitations early in the HLS process is essential to avoid costly design iterations and ensure convergence toward a feasible and deployable FPGA implementation.

Additionally, Super Logic Region (SLR)-aware optimizations are crucial for multi-SLR FPGA architectures. A Super Logic Region (SLR) is a physically distinct section of an FPGA die in Stacked Silicon Interconnect (SSI) technology. Each SLR contains a subset of the device's computational resources, including DSP, LUT and FF, similar to those in monolithic FPGA devices.

Effectively mapping tasks across SLRs is critical for balancing resource utilization and minimizing routing congestion, as inter-SLR communication introduces additional latency and can become a performance bottleneck. By integrating SLR-aware task partitioning, optimized scheduling, and efficient data movement strategies, FPGA designs can achieve higher scalability, improved parallelism, and enhanced synthesis feasibility. Ensuring that computation and memory access patterns align with the physical structure of multi-SLR FPGAs is essential for achieving high QoR while maintaining timing closure and efficient resource allocation.

Some recent frameworks, such as RapidStream-TAPA \cite{10.1145/3609335,10.1145/3593025} and PASTA \cite{pasta}, aim to improve the design scalability and performance of HLS programs on modern multi-die FPGAs through hardware-aware co-optimization of HLS and physical design. While RapidStream-TAPA focuses on latency-insensitive, FIFO-based communication and exploits coarse-grained floorplanning and pipelining for timing closure and parallel compilation, PASTA extends this model to include buffer-based inter-task communication using a generalized channel abstraction. Despite their advances, both frameworks still require a manually optimized task-parallel input code, which demands substantial expertise in HLS and hardware design, limiting their accessibility for non-expert users.

\subsection{Challenges}

While existing HLS optimizations provide substantial performance improvements, several challenges remain. One of the primary limitations is that many prior works restrict their optimization space to specific techniques or separate their optimization processes into multiple independent steps. This fragmented approach can lead to incoherent design decisions, where optimizations applied at one stage may not align with or may even counteract those applied at subsequent stages. Refer to Table \ref{tab:related} for a detailed comparison of these approaches.

A common oversimplification in many frameworks is the assumption that all data resides on-chip e.g., \cite{stream_hls, scalehls}. While this simplifies memory access patterns in design space exploration, it often results in low Quality of Results (QoR) when deployed on real hardware. This assumption prevents frameworks from incorporating crucial tiling strategies, which are essential for optimizing data locality, reducing memory access overhead, and ensuring efficient off-chip communication. When off-chip memory management is treated as a separate optimization step, decisions made in earlier phases may restrict the effectiveness of later memory optimizations, leading to suboptimal performance and resource utilization.

The limitations of current HLS optimization frameworks that rely on shared buffering are exemplified through the \textit{3mm} kernel. It consists of two independent matrix multiplications, whose outputs serve as inputs for a final matrix multiplication. The 3mm kernel computes: G = (A × B) × (C × D) where intermediate matrices E and F store the results of the two first multiplications.

As illustrated in Listing \ref{lst:3mm_motiv}, each matrix multiplication consists of deeply nested loops that are well-suited to parallelization techniques such as loop unrolling, pipelining, and computation-communication overlap. However, conventional shared buffering strategies often fall short in effectively leveraging concurrency, primarily due to their limited exploitation of dataflow principles and insufficient overlap between computation and communication.

Another major challenge is balancing parallelism and routing complexity. Excessive parallelization can lead to routing congestion, increased place-and-route time, and ultimately, failure to generate a valid bitstream. Current methodologies lack adaptive mechanisms to dynamically adjust parallelism levels based on available routing resources and FPGA architecture constraints. Without such mechanisms, achieving high-performance designs often requires manual intervention and iterative fine-tuning.

While pragma-based optimizations such as \textit{unroll} and \textit{pipeline} significantly improve performance, their effectiveness is inherently tied to the underlying code structure. Without proper transformations to expose parallelism and optimize memory access patterns, pragma insertion alone cannot fully unlock the potential of FPGA acceleration.

Addressing these challenges requires a holistic approach that integrates memory-aware optimizations, automated design space exploration, and adaptive scheduling techniques. Future research should focus on frameworks that jointly optimize data placement, computational parallelism, and resource utilization while considering real hardware constraints. By developing more cohesive and intelligent HLS optimization strategies, FPGA-based acceleration can achieve greater efficiency, scalability, and deployment feasibility.



Model-free approaches, such as AutoDSE \cite{autodse} and AI-based frameworks like HARP \cite{harp, ironman, 9803218}, rely on either HLS compilers or AI models to predict performance and resource utilization. However, these methods are currently limited to pragma insertion and space enumeration. Even though AI-based models provide fast estimations (on the order of milliseconds), exhaustively exploring large design spaces remains infeasible. Our previous work, NLP-DSE \cite{nlp_dse}, addresses this limitation by leveraging NLP techniques to efficiently explore pragma configurations. However, like the other approaches, it remains restricted to pragma insertion.

PolyOptHLS \cite{pouchet:fpga13} focuses on optimizing designs by minimizing memory transfers. While reducing communication latency can significantly improve performance, it does not always lead to optimal results. In many cases, the primary bottleneck may shift from communication to computation, limiting overall efficiency and potentially degrading the quality of results (QoR). This approach, therefore, lacks a balanced optimization strategy that considers both computation and communication trade-offs.

ScaleHLS \cite{scalehls} and POM \cite{pom} extend optimization beyond pragma insertion by incorporating code transformations. However, their exploration space is constrained by the assumption that data is already on-chip. Additionally, their transformations are based on heuristics, such as permuting the reduction loop to the outermost level. While they employ a cost model to minimize computation latency, they still rely on exhaustive enumeration of possible configurations.

Allo \cite{allo} aims to reduce development time and enhance design quality through a composable programming model for hardware accelerator design. By decoupling algorithm specification from hardware customization and enabling modular schedule composition, Allo supports holistic dataflow optimization and verifiable transformations for large-scale, multi-kernel designs.
However, it still requires significant manual intervention from an expert, limiting its usability for non-specialists.

Our previous work, Sisyphus \cite{sisyphus}, enables both code transformations and pragma insertion but is restricted to optimizing a single task. It lacks support for optimizations that dataflow techniques can provide, which would be necessary for broader efficiency improvements.

Stream-HLS \cite{stream_hls} effectively leverages dataflow pragmas by selecting a good loop ordering strategy to maximize streaming efficiency. However, it imposes constraints on the design space by assuming that data is on-chip. Moreover, its approach to increasing parallelism relies on multiple FIFOs, which is not generalizable when off-chip memory access is required.

A significant limitation of all these frameworks is that none of them account for hardware constraints in their exploration space. While this simplifies the exploration process, it reduces real-world applicability. Additionally, frameworks such as Merlin-based tools (AutoDSE, HARP, NLP-DSE) and Sisyphus can generate all necessary components, including off-chip memory, for bitstream generation. However, they are constrained to a single Super Logic Region (SLR), leading to under-utilization of the FPGA board. Although generating a design that spans multiple SLRs is theoretically possible, it significantly increases the risk of bitstream generation failures. Even if the bitstream is successfully generated, timing violations often degrade performance.


\begin{figure}[!hbt]
\centering
\begin{minipage}{0.48\linewidth}
\begin{lstlisting}[language=C, label={lst:3mm_motiv}, caption={\textit{Reference Implementation of the 3mm Kernel from PolyBench.}
The 3mm kernel computes: G = (A × B) × (C × D) where intermediate matrices E and F store the results of the two first multiplications.}, escapeinside={(*@}{@*)}]
for (i = 0; i < 180; i++)  // MM 1
  for (j = 0; j < 190; j++) {
    E[i][j] = 0.0; // S0
    for (k = 0; k < 200; ++k) 
      E[i][j] += A[i][k]*B[k][j];}// S1
for (i = 0; i < 190; i++)  // MM 2
  for (j = 0; j < 210; j++) {
    F[i][j] = 0.0; // S2
    for (k = 0; k < 220; ++k) 
      F[i][j] += C[i][k]*D[k][j];}// S3
for (i = 0; i < 180; i++)  // MM 3
  for (j = 0; j < 210; j++) {
    G[i][j] = 0.0; // S4
    for (k = 0; k < 190; ++k) 
      G[i][j] += E[i][k]*F[k][j];}// S5
\end{lstlisting}
\end{minipage}
\hfill
\begin{minipage}{0.48\linewidth}
\begin{lstlisting}[language=C++, label={lst:3mm_motiv2}, caption={{Direct Transformation of Listing~\ref{lst:3mm_motiv} Where Each Loop Body Becomes a Separate Task}}]
for (i = 0; i < 180; i++)  // Task 0
  for (j = 0; j < 190; j++) 
    E[i][j] = 0.0; // S0
for (i = 0; i < 180; i++)  // Task 1
  for (j = 0; j < 190; j++) 
    for (k = 0; k < 200; ++k) 
      E[i][j] += A[i][k]*B[k][j];}// S1
for (i = 0; i < 190; i++)  // Task 2
  for (j = 0; j < 210; j++) 
    F[i][j] = 0.0; // S2
for (i = 0; i < 190; i++)  // Task 3
  for (j = 0; j < 210; j++) 
    for (k = 0; k < 220; ++k) 
      F[i][j] += C[i][k]*D[k][j];}// S3
for (i = 0; i < 180; i++)  // Task 4
  for (j = 0; j < 210; j++) 
    G[i][j] = 0.0; // S4
for (i = 0; i < 180; i++)  // Task 5
  for (j = 0; j < 210; j++) 
    for (k = 0; k < 190; ++k) 
      G[i][j] += E[i][k]*F[k][j];}// S5
\end{lstlisting}
\end{minipage}
\end{figure}

\subsection{Overview of \framework}

We introduce \framework, a unified optimization framework designed to efficiently navigate complex design spaces using a Non-Linear Programming (NLP) cost model. \framework streamlines FPGA design development by providing an intuitive, engineer-friendly interface with tunable parameters, including tile size, array partitioning, and loop unrolling factors.

The framework integrates all the optimization techniques discussed in previous sections, ensuring a comprehensive and cohesive approach to performance enhancement. Additionally, \framework is SLR-aware, enabling intelligent task distribution across Super Logic Regions (SLRs). This capability not only simplifies bitstream generation but also optimizes the trade-off between computational performance and resource utilization.






\begin{figure}[h]
\footnotesize{
\centering
\begin{tikzpicture}[node distance=0.4cm and 1.4cm] 

\node[process, minimum width=0.9cm, xshift=-4cm] (cpp) {\parbox{0.5cm}{\centering  \textbf{C++ Code}}};

\node[process, right=2cm of cpp] (ir) {\parbox{3.5cm}{\centering \textbf{Intermediate Representation} \\
\textit{(dependencies, trip counts)}}};

\node[process, right=of ir] (task) {\parbox{3.5cm}{
    \centering
    \textbf{Task-flow Graph}\\[0.2cm]
    \begin{tikzpicture}[scale=0.6, every node/.style={circleproc}]
      \node (n1) at (0, 0.6) {1};
      \node (n2) at (1.2, 0.6) {2};
      \node (n3) at (0, -0.6) {3};
      \node (n4) at (1.2, -0.6) {4};
      \draw[arrow] (n1) -- (n3);
      \draw[arrow] (n2) -- (n4);
      \draw[arrow] (n3) -- (n4);
    \end{tikzpicture}
}};

\node[process, below=of task] (nlp) {\parbox{3.8cm}{\centering \textbf{Formulate a Non-Linear Problem} \\ encoding the full optimization design space}};

\node[process, fill=black, text=white, below=of nlp] (params) {\textbf{All Parameters of the Space} \\ \textit{e.g.} data-tile of array $A$ is $\texttt{32} \times \texttt{64}$};

\node[process, left=of params] (hlsfiles) {\textbf{ Files for HLS Compiler} \\ \textit{HLS-C++, OpenCL, Config}};


\node[process, left=of hlsfiles, xshift=0cm] (bitstream) {\parbox{2.8cm}{\centering \textbf{Bitstream} \\ with HLS \\ Compiler}};


\draw[arrow] (cpp) -- node[above,xshift=-11pt]{\shortstack{\textit{Extract} \\ \textit{with PoCC}}} (ir);

\draw[arrow] (ir) -- node[above]{\textit{Generate}} (task);
\draw[arrow] (ir) -- (nlp);
\draw[arrow] (task) -- (nlp);
\draw[arrow] (nlp) -- node[right]{\textit{Solve NLP}} (params);
\draw[arrow] (params) -- node[above, sloped]{\shortstack{\textit{Generate} }} (hlsfiles);
\draw[arrow] (hlsfiles) -- (bitstream);

\begin{pgfonlayer}{background}
  \node[draw=prometheusblue, thick, fill=prometheusblue!100, rounded corners=6pt,
        fit=(ir)(task)(nlp)(params)(hlsfiles), inner sep=0.5cm] (groupbox) {};
  \node[header, anchor=north west] at (groupbox.north west) {Prometheus};
\end{pgfonlayer}

\end{tikzpicture}
}
\caption{\textit{Overview of the Prometheus Framework Workflow.} This diagram illustrates the end-to-end pipeline of the Prometheus optimization framework, starting from C++ source code and proceeding through intermediate representation extraction, task-flow graph generation, Non-Linear Programming (NLP)-based design space exploration, and final compilation into FPGA bitstreams using HLS compilers.}
\label{fig:flow}

\end{figure}

\begin{table}[!htb]
\centering
\renewcommand{\arraystretch}{1.4}
\setlength{\tabcolsep}{8pt}
\footnotesize{
\begin{tabular}{|l|p{3cm}|p{6.2cm}|}
\hline
\textbf{Group} & \textbf{Parameter} & \textbf{Description} \\
\hline
\multicolumn{3}{|c|}{\textbf{Design Variables}} \\
\hline

\multirow{5}{*}{\textbf{Memory}} 
  & Bit Width ($BW_a$) & Width of memory transfer in bits; \\
%
  & Data-tile Size (modeled via Transfer Level $t_{a,l}$ in Section~\ref{sec:nlp}) & Size of the data tile of array $a$ transferred either from off-chip to on-chip memory or between tasks. The transfer level determines this size. \\
& Data-tile Reuse Size (modeled via Reuse Level $d_{a,l}$ in Section~\ref{sec:nlp}) & Size of the data tile of array $a$ reused within a task. The reuse level dictates this tile size. \\
  & Buffering ($N_a$) & Number of buffers (2 for double buffering, 3 if read \& write), enabling overlap of load/compute/store. \\
  & Communication Padding & Additional data added to enable wider burst transfers and better bandwidth utilization. \\
\hline

\multirow{3}{*}{\textbf{Parallelism}} 
& Unroll Factors ($TC^l_{\text{intra}}$) & Unroll factor for each task, determined through tiling, where the intra-tile loop is fully unrolled to expose fine-grained parallelism. \\
  & Array Partitioning ($AP_{a,d}$) & Number of partitions for array $a$ along dimension $d$ to support parallel accesses. \\
  & Compute Padding & Padding applied to loop trip counts to increase legal unroll factors. \\
\hline

\multirow{3}{*}{\textbf{Code Structure}} 
  & Loop Permutations & Legal reordering of loops to optimize data locality and parallelism; synchronized across fused statements. \\
  & Tiling & Splitting loops into inter- and intra-tile levels for optimization and pipelining. \\
  & SLR Assignment ($slr_t$) & Mapping of each task $t$ to a Super Logic Region (SLR) to enable spatial task distribution. \\
\hline

\multicolumn{3}{|c|}{\textbf{Design Constraints}} \\
\hline

\multirow{3}{*}{\textbf{Resource Limits}} 
  & Max Array Partitioning  & Total partitions across all arrays must not exceed architectural limits. \\
  & DSP Budget & Maximum DSP slices per SLR; used to limit total DSP usage by pipelined/unrolled tasks. \\
  & On-Chip Memory Available & Total memory usage by buffered tiles (considering reuse and buffering) must be within BRAM capacity. \\
\hline
\end{tabular}
}
\caption{\textit{Design Variables and Architectural Constraints in the Prometheus Optimization Space.} 
This table outlines the key configurable parameters used in Prometheus—including loop tiling, unrolling, array partitioning, and buffer management—along with hardware constraints such as resource limits and memory capacity. 
The cost model also accounts for concurrent task execution enabled by the \texttt{dataflow} pragma, as well as computation-communication overlap achieved through automatic double or triple buffering. 
These elements jointly define the valid and efficient design space explored by Prometheus.}
\label{tab:nlp_design_space}
\end{table}

Figure \ref{fig:flow} illustrates the workflow of \framework. The process begins with an input C++ code, from which we extract an intermediate representation containing dependencies, trip counts, schedules, and other relevant information. This extraction is performed using PoCC \cite{pocc-web}, which provides all necessary details for analysis.
Next, we construct the task-flow graph based on this extracted information and formulate the corresponding Non-Linear Programming (NLP) problem. Solving this NLP problem determines the theoretical optimal parameters for the design space.
Once the NLP solver provides these parameters, we have all the required information to proceed. At this stage, we automatically generate the HLS-C++ code along with all the necessary files to produce the FPGA bitstream, ensuring an efficient and fully automated compilation process.

To support efficient exploration of the design space, Prometheus defines a set of tunable parameters and architectural constraints that capture the key aspects of FPGA acceleration. Table~\ref{tab:nlp_design_space} provides a detailed overview of these parameters, including memory-related configurations (e.g., bitwidth, tiling, and reuse levels), parallelism controls (e.g., loop unrolling and array partitioning), and structural transformations (e.g., loop permutations and SLR assignments). In addition, the table outlines hardware constraints such as DSP budgets, on-chip memory limits, and maximum legal array partitioning. These variables and constraints jointly define the feasible design space that the NLP-based optimization engine systematically explores to generate high-performance, hardware-aware FPGA implementations.

The pseudo-code in Listing \ref{lst:3mm_prometheus} demonstrates how \framework processes the \textit{3mm} kernel (Listing \ref{lst:3mm_motiv}). Load and store operations manage data transfers to and from off-chip memory, while send and receive operations handle inter-task communication using FIFO. 

Each task\textit{i} in the pseudo-code corresponds to the fully unrolled computation of an intra-tile for statement S\textit{i} in Listing \ref{lst:3mm_motiv}. An example of such a task is illustrated in Listing \ref{lst:3mm_task3}. 

Arrays \(E\), \(F\), and \(G\) are initialized to zero within their respective tasks (e.g., S0, S2, and S4) and are not preloaded, reducing unnecessary memory overhead.

\framework enhances computation efficiency by fusing statements that produce the same outputs (e.g., in Listing \ref{lst:3mm_motiv}), and it automatically formulates an NLP problem to determine the theoretically optimal parameters, such as loop schedules, array bit widths (e.g., 512 bits), tile sizes, reuse buffer sizes, transfer locations, and padding. The framework dynamically adjusts bit widths for arrays like \(F\), \(D\), and \(G\) to achieve a better balance between parallelism and resource usage.

\begin{figure}[!htb]
\begin{lstlisting}[label={lst:3mm_prometheus},caption={\textit{Transformed and Fused 3mm Kernel Pseudocode Generated by Prometheus.} This code highlights the result of Prometheus optimizations, including fused task generation, tiled loops, memory buffer management, and pipelined execution with dataflow pragmas.},escapeinside={(*@}{@*)}]
/*************** Fused Task 0 ****************/
float B[204][192]; load_B(B);
for (i0 = 0; i0 < 18; i0++)//inter-tile loop
  float A[10][204]; load_A(A);
  for (j0 = 0; j0 < 6; j0++){//inter-tile loop
    float E[10][32]; 
    task0(E); // S0 + intra-tile loops 
    for (k0 = 0; k0 < 51; ++k0)//inter-tile loop
#pragma HLS pipeline II=3
      task1(E, A, B); // S1 + intra-tile loops 
    store_E(E); sent_E(E);
/*************** Fused Task 1 ****************/
float C[190][222]; load_C(C);
for (j0 = 0; j0 < 7; j0++)//inter-tile loop
  float D[222][32]; load_D(D);
  for (i0 = 0; i0 < 10; i0++){//inter-tile loop
    float F[19][32]; 
    task2(F); // S2 + intra-tile loops 
    for (k0 = 0; k0 < 74; ++k0)//inter-tile loop
#pragma HLS pipeline II=3
      task3(F, C, D); // S3 + intra-tile loops 
    store_F(F); sent_F(F);
/*************** Fused Task 2 ****************/
float E[180][192];
for (j0 = 0; j0 < 7; j0++)//inter-tile loop
   float F[192][32];
   receive_F(F);
  for (i0 = 0; i0 < 18; i0++){//inter-tile loop
    float G[10][32]; 
    receive_E(E);
    task4(G); // S4 + intra-tile loops 
    for (k0 = 0; k0 < 32; ++k0)//inter-tile loop
#pragma HLS pipeline II=3
      task5(G, E, F); // S5 + intra-tile loops 
    store_G(G);
\end{lstlisting}
\end{figure}

Efficient computation-communication overlap is achieved through ping-pong buffering, while concurrent task execution and FIFO-triggered dependent tasks maximize overall performance.
Fused Tasks 0 and 1 execute concurrently, while Fused Task 2 begins as soon as the data tiles for \(F\) and \(E\) become available.

To assess the effectiveness of our proposed framework, we evaluate the performance of the 3mm kernel using several state-of-the-art HLS-based FPGA optimization tools. As shown in Table~\ref{tab:3mm_horizontal}, Prometheus significantly outperforms existing frameworks in terms of throughput, achieving 368.36 GF/s. This represents a substantial improvement over Sisyphus (178.97 GF/s) and Stream-HLS (174.00 GF/s), and far exceeds the results of Allo, ScaleHLS, and AutoDSE. The superior performance of Prometheus stems from its holistic design space exploration strategy, which integrates loop transformations, memory tiling, concurrent task execution, and SLR-aware scheduling to deliver optimized and hardware-feasible designs.

\begin{table*}[h]
\centering

\footnotesize{
\begin{tabular}{lcccccc}
\toprule
\textbf{Metric} & \textbf{Prometheus} & \textbf{Sisyphus} & \textbf{Stream-HLS} & \textbf{Allo} & \textbf{ScaleHLS} & \textbf{AutoDSE} \\
\midrule
Throughput (GF/s) & \textbf{368.36} & 178.97 & 174.00 & 60.40 & 43.04 & 1.74 \\
\bottomrule
\end{tabular}
}
\caption{\textit{Measured Throughput of the 3mm Kernel Using Various FPGA Frameworks.}
This table presents the runtime throughput (in GF/s) obtained from RTL simulation of the 3mm kernel across multiple frameworks. Prometheus is shown to outperform other methods, including Sisyphus, Stream-HLS, and AutoDSE.}
\label{tab:3mm_horizontal}
\end{table*}


\section{Code Transformation}
\label{sec:code_transformation}

In our dataflow model, we adopt a synchronous dataflow where the sizes of the arrays are known during compile time. This compile-time awareness enables us to construct a precise model that facilitates rigorous optimizations.
To leverage FPGA parallelism, we implement an acyclic dataflow graph, ensuring parent nodes do not receive data from their children. While this constraint limits graph configurations and may increase resource usage, it reduces overall latency. To support this structure, we inline \cite{241594} each function in the input code to generate the required acyclic graph.
Our primary objective is to minimize latency within resource constraints by overlapping communication and computation within tasks and executing independent tasks concurrently. To achieve this, we apply various transformations and optimizations, explored in this section, and navigate the design space using an NLP-based approach, detailed in Section \ref{sec:nlp}.

\subsection{Dependency Graph Creation}
Our process starts with affine C/C++ code as input, which undergoes maximal distribution to ensure each loop body contains only one statement, provided no dependencies exist within the loop. ISCC \cite{iscc} verifies the legality of these transformations, ensuring dependencies are preserved.
After achieving full distribution, 
we construct a dependency graph using PoCC \cite{pocc-web}. PoCC provides the necessary information about the schedule and dependencies, enabling us to build the graph. In this graph, the nodes represent tasks, while the edges capture data communication arising from inter-task dependencies.
Tasks with identical outputs are then merged (when legal), creating fused tasks with output-stationary properties. 
 This ensures that each tile’s output is handled (loaded, computed, and either stored or transmitted) only once.
If a dependency prevents distribution, the framework will still function but with a more limited optimization space, making it more effective when distribution is possible.

\newtext{Figure~\ref{fig:ex_graph} shows the task graph of the \textit{3mm} kernel (Listing~\ref{lst:3mm_motiv2}), where each loop body corresponds to a distinct computation task.}

\begin{figure}[H]
    \centering
    \begin{tikzpicture}[scale=0.009, every node/.style={circle, draw, minimum size=5mm, inner sep=0pt}]
        \node (0) at (120, 260) {T0};
        \node (1) at (200, 260) {T1};
        \node (2) at (120, 200) {T2};
        \node (3) at (200, 200)  {T3};
        \node (4) at (400, 230) {T4};
        \node (5) at (300, 230) {T5};

        \draw[->] (2) -- (3);
        \draw[->] (0) -- (1);
        \draw[->] (4) -- (5);
        \draw[->] (1) -- (5);
        \draw[->] (3) -- (5);
    \end{tikzpicture}
    \caption{Dataflow graph of the \texttt{3mm} kernel. Nodes represent computation tasks; edges denote data dependencies.}
    \label{fig:ex_graph}
\end{figure}

\noindent
\framework's solution space includes:

\subsection{Data-tiling and Padding}  
In the fused dependency graph, edges represent data communication between tasks as well as between tasks and off-chip memory, involving the transfer of data tiles with specified sizes.
Data tiling divides loop iterations into smaller tiles, splitting array accesses into subsets. Each loop $l$ iterating over an array $a$ is divided into an outer loop ($TC^l_{inter}$) and an inner loop ($TC^l_{intra}$), with feasible permutations applied. 
For each array and dimension iterated by the loop $l$, the tile factor is a common choice that influences all arrays iterated by this loop.
%

To optimize memory transfers, padding is applied to arrays in two ways. Simple padding increases the bit width ($BW_a$) for efficient transfers while maintaining the original loop trip count. Composite padding adjusts both $BW_a$ and the loop trip count ($TC^l$) to support unroll factors that do not evenly divide the original trip count, allowing irregular tile sizes and expanding the design space.
Tile sizes are consistent within a fused task but vary between tasks. In Listing \ref{lst:3mm_prometheus}, the tile size of array F is $19 \times 32$ in Fused Task 1 and $192 \times 32$ in Fused Task 2.

\subsection{Fine-grained Parallelism}
Data-tile selection involves splitting each loop and permuting them to create two levels of the original loops. Using ISCC \cite{iscc}, we verify the legality of these permutations, resulting in two loop levels: inter-tile (outer) and intra-tile (inner). If permutation is not feasible, the inter-tile loop retains its legal position.
%
For tasks belonging to the same fused task, we merge their inter-tile loops, which are non-reduction. For instance, in Listing \ref{lst:3mm_prometheus}, the inter-tile loops on lines 3 and 5 iterate over task0 and task1.
The intra-tile loops are fully unrolled, ensuring that all data accesses within the intra-tile remain on-chip.
Array partitioning, determined by the unroll factor, ensures data resides in separate BRAM banks for simultaneous access. Reduction loops across inter-tiles are pipelined with an initiation interval ($II = n > 1$), where  n  matches the reduction latency. For instance, in Listing \ref{lst:3mm_prometheus}, additions take 3 cycles, resulting in $II = 3$.
%


\subsection{Loop Order}
Due to the full unrolling of the intra-task, there is no need to select the loop order within the intra-tile. 
We place the inter-tile reduction loops directly above the task and are pipelined.
If multiple reduction loops exist, we rank them by the size of their trip counts, placing the loop with the highest trip count innermost to ensure an efficient pipeline.
This setup provides the flexibility to choose the order of the inter-tile loops that are not reduction loops. This order will subsequently be determined by the NLP (cf. Section \ref{sec:nlp}).
As shown in Listing \ref{lst:3mm_prometheus}, the loop order of Fused Tasks 1 and 2 is permuted compared to the original program in Listing \ref{lst:3mm_motiv}.

\subsection{Automatic Overlapping of Communication and Computation}
To overlap communication and computation, we use on-chip buffers. The buffer size and data transfer location are determined by various options explored via NLP (cf. Section \ref{sec:nlp}). Since data must stay on-chip for intra-task computation, transfers occur either below an inter-tile loop or before any loops start. 
The transfer position determines the data tile size, covering all data accessed below it.
To improve data reuse, buffer size can match or exceed the transferred data tile. Similar to transfer location, buffer size is determined by its position relative to inter-tile loops or before any loops. Two boolean variables, $d_{a,l}$ and $t_{a,l}$, define and transfer array $a$ under loop $l$ when set to true. If defined or transferred before loops, $l=0$.
Double-buffering is used for read-only or write-only arrays, and triple-buffering for arrays that are both read and written. Following \cite{DBLP:journals/corr/abs-1807-01340}, we perform an initial load, then overlap loading the next tile with computing the current one. 


\subsection{Coarse-Grained Parallelism: Automatic Execution of Concurrent Tasks}
Due to the use of the dataflow pragma, tasks can begin computation as soon as they have sufficient data, allowing for concurrent execution. This concurrency significantly increases overall performance.

\subsection{Memory Transfer}
 Communication latency is reduced by transferring more data per cycle using increased bit width with padding. Read-only arrays are duplicated in off-chip memory for tasks with multiple reads, eliminating feed-through logic. For non-read-only arrays, data passes between tasks via FIFOs, using the same bit width as for off-chip memory transfer.

\section{NLP Formulation}
\label{sec:nlp}

To determine the tile size, loop order, bit width, and memory transfers, we formulate a cost model as a NLP problem aimed at minimizing overall latency. 
This approach builds upon the methodology proposed in our previous work \cite{nlp_dse, nlp_dse_poster, sisyphus}, which we have further adapted to meet the specific requirements and constraints of our current framework. In our work, we incorporate dataflow considerations along with all the optimizations detailed in Section \ref{sec:code_transformation}.
We employ PoCC \cite{pocc-web} to extract compile-time information such as schedules, loop trip counts, dependencies, and operation counts per statement. ISCC \cite{iscc} then generates all legal permutations for each loop body.

Table \ref{tab:constant2} 
delineates the sets, variables, and constants utilized in our NLP formulation.



\begin{table}[!htb]
\centering
\footnotesize{
\begin{tabular}{@{}p{0.12\linewidth}p{0.84\linewidth}@{}}
\toprule

Constants & Description \\
\midrule

$II_l$ & II of the loop $l$ \\
$IL_{par}$, 

$IL_{red}$ & Iteration Latency of the operations without ($par$) and with ($red$) dependencies of the statement $s$\\
$TC^l_{ori}$ & Original Trip Count of the loop $l$ \\
$f_{a,l}$ & Footprint of the array $a$ if transferred to on-chip after the loop $l$ \\
$N_{FT}$ & Number of fused task \\
$DSP_{s_{op}}$ & Number of DSP used by the statement $s$ for the operation $op$\\
$DSP$ & Number of DSP available for the FPGA used \\
$max_{part}$ & Maximum array partitioning \\
$SLR$ & Number of SLR available for the FPGA used \\
\midrule

Variables & Description \\
\midrule

$TC^{l}_{intra}$,

$TC^{l}_{inter}$ & TC of the loop $l$ for the intra and inter tile \\
$TC^l$ & Trip Count of the loop $l$ after padding \\
$S^{last}_{a}$ & Size of the last dimension of the array $a$ transferred on-chip \\
$BW_a$ & Bit width of the array $a$ \\
$t_{a,l},d_{a,l}$ & Boolean to know if the array $a$ is transferred and defined (respectively) on-chip after the loop $l$ in the inter-tile\\
$p_i^{l}$ & Position 
of the loop $l$
under the $i$-th permutation \\
$slr_t$ & ID of the SLR use by the task $t$ \\


\midrule
Sets & Description \\
\midrule
$\mathcal{L},\mathcal{A},\mathcal{S}$ & The set of loops, arrays and statements \\ 
$\mathcal{L}_s$ &  The set of loops which iterate the statement $s$ \\
$\mathcal{L}_s^{red}$ &  The set of reduction loops which iterate the statement $s$ \\
$\mathcal{L}_a^{last}$ &  The set of loop which iterate the last dimension of the array $a$\\
$\mathcal{L}_{inter},$ 

$\mathcal{L}_{intra}$ &  The set of  loops which belong to the inter-tile and intra-tile respectively \\
$\mathcal{B}$ & Set of possible burst size for the data type \\
$\mathcal{C}_{a_d}$ & The set of loops which iterates the array $a$ at the dimension $d$ \\ 
$AP_{a,d}$ & Array Partition for the array $a$ in dimension $d$ \\
$\mathcal{P}_{s}$ & All permutation of the loops which iterate the statement $s$ \\
$\mathcal{F}_{i}$ & The set of statements which belong to the fused task $i$ \\
$\mathcal{T}$ & The set of tasks in the dataflow \\
\bottomrule
\vspace{0.0005cm}
\end{tabular}%
}
\caption{\textit{Mathematical Notation for Constants, Variables, and Sets in the NLP-Based Optimization Model.}
This table defines the formal notation used in Prometheus’ Non-Linear Programming (NLP) model for design space exploration, including loop trip counts, bitwidths, SLR mappings, resource limits, and legal transformations.}
\label{tab:constant2}
\end{table}

\subsection{Constraints}

We now describe the constraints by using the code of Listings \ref{lst:3mm_motiv}, \ref{lst:3mm_prometheus} and \ref{lst:3mm_task3}.

\begin{figure}

\begin{lstlisting}[label={lst:3mm_task3},caption={\textit{Implementation of Task 3 in the 3mm Kernel (from Listing \ref{lst:3mm_prometheus}).}
This listing illustrates the structure of Task 3, where the intra-tile computation is fully unrolled to expose fine-grained parallelism. The unrolling enables concurrent execution of loop iterations.},escapeinside={(*@}{@*)}]
for (int j1 = 0; j1 < 32; j1++) 
#pragma HLS unroll
  for (int i1 = 0; i1 < 19; i1++) 
#pragma HLS unroll
    for (int k1 = 0; k1 < 3; k1++) {
#pragma HLS unroll
      j=j0*32+j1; i=i0*19+i1; k=k0*3+k1;
      F[i1][j1] += C[i][k] * D[k][j1];}
\end{lstlisting}
\end{figure}

\subsubsection{Data-tiling and Unroll Factor}
The intra-tile transformation, as explained in Section \ref{sec:code_transformation}, can divide either the original loop trip count or the original trip count with padding, thereby increasing the range of possibilities. Equation \ref{eq:uf} ensures that the trip count of the intra-tile is a divisor of one of these two possibilities.
The user has the option to constrain the padding using Equation \ref{eq:cons_pad}, which simplifies the solution space for the NLP solver.
For instance, in Listing \ref{lst:3mm_task3}, for the array \( F \), the loop \( j \) (line 7 in Listing \ref{lst:3mm_motiv}) has been split into \( j0 \) (line 14 in Listing \ref{lst:3mm_prometheus}) and \( j1 \) (intra-tile).  
The trip count of the intra-tile loop \( j1 \), denoted as \( TC^{j1}_{\text{intra}} = 32 \), does not evenly divide the original trip count \( TC^{j}_{\text{ori}} = 210 \), but it does divide the trip count of the padded loop \( TC^{j} = 224 \).

\begin{eqnarray}
\label{eq:uf}
\forall l \in \mathcal{L}, TC^l_{intra} \% TC^l_{ori} == 0 || TC^l_{intra} \% TC^l ==0 \\
\label{eq:cons_pad}
(opt) \forall l \in \mathcal{L}, \exists n \in \mathbb{N} \leq N \in \mathbb{N}, \text{ s.t. } TC^l = TC^l_{ori} + n
\end{eqnarray}

\subsubsection{Bit Width}
$\mathcal{B}$ denotes the number of elements that can be transferred simultaneously, determined by the bit width and the data type. Hence, if we have a bit width under 512 bits for  \textit{float} the set is $\{1,2,4,8,16\}$.
Equation \ref{eq:dt_tile} computes the bit width for each array based on the last dimension of the data-tile transferred on-chip.
For example, the array D in Fused task 1 is transferred in line 15 with a size of $222 \times 32$, so $S^{last}_{D} = 32$, which is divisible by 16. Therefore, it has a bit width of 16. 

\begin{eqnarray}
\label{eq:dt_tile}
\forall a \in \mathcal{A}, \forall l \in \mathcal{L}_a^{last}, 
\label{eq:dt_tile}
\max_{b \in \mathcal{B}}BW_a =  b \text { s.t. } S^{last}_{a} \% b = 0 
\end{eqnarray}

\subsubsection{Permutation}
Equation \ref{eq:perm_sim} requires the NLP to choose identical permutations for loops that are shared by statements fused within the same task.
For example, in Fused task 1, the loops iterating statement S2 can be permuted as ($i0$, $j0$) or ($j0$, $i0$), and similarly, loops iterating statement S3 can be permuted as ($i0$, $j0$) or ($j0$, $i0$). However, since the loops $i0$ and $j0$ iterate both S2 and S3, they must use the same permutation; either ($i0$, $j0$) for both or ($j0$, $i0$) for both.

\begin{eqnarray}
\label{eq:perm}
\nonumber
\forall i \in \llbracket 0, N_{FT}  \llbracket, \forall (ft_0, ft_1) \in \mathcal{F}_i^2,\\
\label{eq:perm_sim}
\forall (i_0, i_1) \in \mathcal{P}_{ft_0} \times \mathcal{P}_{ft_1}, \forall l \in \mathcal{L}, p_{i0}^l = p_{i1}^l
\end{eqnarray}

\subsubsection{Transfer and Reuse}
Equation \ref{eq:tranfer} 
permits the selection of a single level where each array can be defined (and reused) and transferred.
Equation \ref{eq:transfer_after_reuse} constrains that the definition of the array must occur lexicographically before or at the same time as the transfer. 

The array $E$, defined on line 24 in Listing \ref{lst:3mm_prometheus}, is defined before any loops, so $d_{E,0}=1$ (0 indicating it is defined before any loops). However, it is transferred under the loop $i0$, so $t_{E,i0}=1$. Equation \ref{eq:transfer_after_reuse} simply means that the definition of array E should occur before or at the same level as the transfer. We cannot transfer E under loop $i0$ if E is defined under $k0$.
Similarly, in Listing~\ref{lst:3mm_prometheus}, the array \( A \) is defined and transferred in line 4, with \( d_{A,i0} = 1 \) and \( t_{A,i0} = 1 \).

\begin{eqnarray}
\label{eq:tranfer}
\forall a \in \mathcal{A}, \sum l \in \mathcal{L}_{inter}, t_{a,l} = 1, \sum l \in \mathcal{L}_{inter}, d_{a,l} = 1\\
\label{eq:transfer_after_reuse}
\forall a \in \mathcal{A}, \sum (l_0, l_1) \in \mathcal{L}_{inter}^2 | 
d_{a,l_0},t_{a,l_1}=1, 
\text{ then } l_0  \preccurlyeq l_1 
\end{eqnarray}

\subsubsection{On-chip Memory} 
Equation \ref{eq:tot_foot} constrains the footprint of the array to be within the available resources, based on where the array is defined, the number of double buffers used and the footprint of the array transferred at this level.

\begin{eqnarray}
\label{eq:tot_foot}
 \sum_{a \in \mathcal{A}} \sum_{l \in \mathcal{L}} 
d_{a,l}   \times f_{a,l} \times N_{a} \leq Mem, 
\end{eqnarray}
with $N_{a}$ being the number of double buffer for the array $a$.

\subsubsection{Array Partitioning}
Equation \ref{eq:max_part1} limits the maximum partitioning of each array. This partitioning is crucial as it impacts the maximum unroll factor, necessitating the distribution of data across different BRAM banks under fully unrolled loops, thereby influencing the utilization of BRAMs. Equation \ref{eq:max_part2} computes the array partitioning needed for each array based on the trip count of the fully unrolled intra-tile loops.

Array \texttt{D} in Listing~\ref{lst:3mm_task3} is traversed by two unrolled loops: \( k1 \), which iterates 3 times (\( AP_{D,0} = 3 \)), and \( j1 \), which iterates 32 times (\( AP_{D,1} = 32 \)). Therefore, the total number of partitions needed is $3 \times 32 = 96$. Consequently, these 96 values of F are stored in different banks, allowing all of them to be accessed in parallel. However, this value must be less than or equal to $max_{part}$.

\begin{eqnarray}
\label{eq:max_part1}
\forall a \in \mathcal{A}, \prod_{d \in \mathbb{N}} 
AP_{a,d} \leq max_{part} \\
\label{eq:max_part2}
\forall a \in \mathcal{A}, \forall d \in \mathbb{N}, \forall l \in C_{a_d}, 
AP_{a,d} = TC^{l}_{intra} == 0 
\end{eqnarray}

\subsubsection{DSP Utilization}
Equation \ref{eq:dsp_pessimist} constrains the number of DSPs used based on the available DSP resources. 
In contrast to \cite{nlp_dse, nlp_dse_poster, sisyphus}, we utilize pessimistic DSP utilization. This approach is necessary because concurrent execution and resource reuse between tasks are not feasible when two tasks can run simultaneously.
%
Given \( DSP_{+} = 2 \), \( DSP_{*} = 3 \), and \( II_{S3} = 3 \), the DSP usage for Task 3 is calculated as \( (2 + 3) \times 1824 \), accounting for the unroll factor. However, since the loop is pipelined with \( II = 3 \), the HLS compiler optimizes resource usage, effectively reducing the DSP count by approximately dividing by \( II \).

\begin{eqnarray}
\label{eq:dsp_pessimist}
     \sum_{op \in \{+,-,*,/ \}} \sum_{s \in \mathcal{S}}
     (DSP_{s_{op}} / II_s) \times \prod_{l \in \mathcal{L_{s}}} TC^{l}_{intra} \leq DSP
\end{eqnarray}

\subsubsection{SLR Selection}
For each task, the NLP determines the SLR on which the task will be implemented (Equation \ref{eq:slr}). Additionally, Equations \ref{eq:tot_foot} and \ref{eq:dsp_pessimist} are applied to each SLR to manage resource allocation per SLR effectively.

\begin{eqnarray}
\label{eq:slr}
     \forall {t \in \mathcal{T}}, slr_t \in \llbracket 0, SLR \llbracket
\end{eqnarray}

\subsection{Objective Function}

The objective of our framework is to minimize the total latency of the design, which is modeled as a directed acyclic graph (DAG). In this dataflow graph, each node represents a computation task (or fused task), and edges represent data dependencies between tasks.

To compute the overall latency, we define the latency of each task \( T \in \mathcal{T} \) as the sum of:
\begin{itemize}
    \item the time at which it becomes ready to start (determined by its dependencies)
    \item and the duration of the task itself
\end{itemize}

We denote:
\begin{itemize}
\item \( \text{Lat}(T) \): the global latency contribution of task \( T \), i.e., the time at which \( T \) finishes,
\item \( \text{Lat}_{\text{task}}(T) \): the execution time (duration) of task \( T \),
\item \( \text{pred}(T) \): the set of immediate predecessor tasks of \( T \),
\item \( \text{shift}_{T_i,T} \): the number of cycles after which \( T \) can start once \( T_i \) has begun (e.g., due to pipelined data production).
\end{itemize}

Then the latency of each task is recursively defined as:

\begin{equation}
\label{eq:dag_latency}
\text{Lat}(T) = \max_{T_i \in \text{pred}(T)} \left[ \text{Lat}(T_i) + \text{shift}_{T_i,T} \right] + \text{Lat}_{\text{task}}(T)
\end{equation}

The overall latency of the design is defined as the latest finish time among all sink tasks (i.e., tasks without successors):

\begin{equation}
\text{Lat}_{\text{total}} = \max_{T \in \mathcal{S}} \left[ \text{Lat}(T) \right]
\end{equation}

where \( \mathcal{S} \subseteq \mathcal{T} \) is the set of sink tasks in the graph.



For the graph in Figure~\ref{fig:ex_graph}, the latency can be expressed step-by-step as:

\begin{equation*}
\left.
\begin{aligned}
& \text{Lat}(T_1) = \text{Lat}(T_0) + \text{shift}_{T_0, T_1} + \text{Lat}_{\text{task}}(T_1) \\
& \text{Lat}(T_3) = \text{Lat}(T_2) + \text{shift}_{T_2, T_3} + \text{Lat}_{\text{task}}(T_3) \\
& \text{Lat}(T_5) = \max\left(
    \text{Lat}(T_1) + \text{shift}_{T_1, T_5},\;
    \text{Lat}(T_3) + \text{shift}_{T_3, T_5},\;
    \text{Lat}(T_4) + \text{shift}_{T_4, T_5}
\right) + \text{Lat}_{\text{task}}(T_5)
\end{aligned}
\right.
\end{equation*}

In the next section, we explain how to compute \(\text{Lat}_{\text{task}}(T)\) for each task using intra-task and inter-tile modeling.

\subsubsection{Intra-Task and Multi-Level Latency Modeling}

Each task \( T \) is composed of computation over a tile of data and may be enclosed in multiple inter-tile loop levels (e.g., tiling for cache reuse). At each level, we must account for both computation and communication (i.e., data transfers).

We model the execution as a combination of:
\begin{itemize}
    \item \textbf{Load:} reading data from off-chip memory (or previous task) into on-chip buffers,
    \item \textbf{Compute:} performing the actual tile computation,
    \item \textbf{Store:} writing the result back to off-chip memory (or to a subsequent task).
\end{itemize}

To enable communication/computation overlap, we use double- or triple-buffering depending on the number of streams. At each inter-tile level \( n \), the total latency is computed as the maximum of the three components, accounting for pipeline shifts and buffer reuse.

\paragraph{Level-Based Latency Recursion.}
Let:
\begin{itemize}
    \item \( \text{Lat}_{n+1} \): latency of executing the inner level \( n+1 \),
    \item \( f_{a,n} \): number of bytes transferred for array \( a \) at level \( n \),
    \item \( BS_a \): memory bandwidth for array \( a \),
    \item \( \alpha \): overlap factor (1 = load or store only, 2 = both).
\end{itemize}

Then, the latency at level \( n \) is:

\begin{equation}
\label{eq:intertile}
\text{Lat}_{n} = \max \left( \text{Lat}_{n+1}, \frac{f_{a,n}}{BS_a} \right) + \text{Lat}_{n+1} + \alpha \cdot \frac{f_{a,n}}{BS_a}
\end{equation}

This formula captures both the initialization overhead and the overlap between communication and computation.

\paragraph{Base Case: Intra-Task Latency.}

The base latency \( \text{Lat}_{n+1} \) is the latency of the innermost computation tile, computed as in~\cite{nlp_dse, sisyphus}.

\begin{equation}
\label{eq:intra}
\text{Lat}_{\text{intra}} = IL_{\text{par}} + IL_{\text{seq}} \cdot \log_2 \left( \prod_{l \in \mathcal{L}^{\text{red}}_s} TC^l_{\text{intra}} \right)
\end{equation}

If the computation is invoked across multiple tiles in a pipelined loop, the full latency becomes:

\begin{equation}
\label{eq:pip_intra}
\text{Lat}_{\text{task}} = \text{Lat}_{\text{intra}} + II \cdot \left( \prod_{l \in \mathcal{L}^{\text{red}}_s} TC^l_{\text{inter}} - 1 \right)
\end{equation}

\paragraph{Final Task Latency.}
The full latency of a task \( T \), denoted \( \text{Lat}_{\text{task}}(T) \), is the latency at the outermost loop level \( n=0 \), computed recursively using Equation~\ref{eq:intertile}, where the base is given by Equation~\ref{eq:pip_intra}.



\section{Code Generation}
\label{sec:code_generation}

\framework takes as input affine C/C++ code and automatically produces an HLS-C++ file, OpenCL host files, and all necessary files for code verification, RTL simulation, and bitstream generation.
The NLP described in Section \ref{sec:nlp} gives the parameters of the space such as loop order, tiling factor, etc. However, in order to be efficient, the code generation needs some specific rules.


\subsection{Communication with Off-Chip Memory}
%
%
To enable overlapping communication and computation, a \textit{load} function transfers data on-chip using FIFOs, ensuring effective overlap management by the compiler.
Furthermore, to guarantee that the \textit{load} operation transfers data with the correct bit width, we automatically restructure the data in off-chip memory to enable sequential loading. Once data accumulates in the FIFO, a \textit{read} function moves it to the shared data-tile buffer, whose size is determined by the NLP.
For outputs, a \textit{write} function transfers data from the buffer to a FIFO, and a \textit{store} function writes it back to off-chip memory.

\newtext{Listing~\ref{lst:3mm_load} shows the load and read functions, which first transfer data from off-chip memory to an on-chip FIFO, and then read from the FIFO into the local buffer allocated for Task 1.}

\begin{figure}[!htb]
\begin{lstlisting}[label={lst:3mm_load},caption={Memory transfer for array \texttt{A} in the 3mm example (Listing~\ref{lst:3mm_motiv}). The data is first loaded from off-chip memory into a FIFO, then read from the FIFO to populate the on-chip buffer},escapeinside={(*@}{@*)}]
void load_vA_for_task1(hls::stream<float16> &fifo_A_from_off_chip_to_S1,
                       float16 vA[2340]) {
#pragma HLS inline off
  for (int i = 0; i < 2340; i++) {
#pragma HLS pipeline II = 1
    fifo_A_from_off_chip_to_S1.write(vA[i]);
  }
}

void read_A_FT0(float A[10][204],
                hls::stream<float16> &fifo_A_from_off_chip_to_S1, int i0) {
#pragma HLS inline off
  if (i0 >= 18) {
    return;
  }
  for (int d0 = 0; d0 < 10; d0++) {
    for (int d1 = 0; d1 < 204; d1 += 16) {
#pragma HLS pipeline II = 1
      float16 tmp_fifo = fifo_A_from_off_chip_to_S1.read();
      for (int j = 0; j < 16; j++){
#pragma HLS unroll
        if (d1 + j < 204)
            A[d0][d1 + j] = tmp_fifo[j];
      }
    }
  }
}
\end{lstlisting}
\end{figure}

\subsection{Communication between the Fused Task}
The communication within the same fused task is not required as they use  shared buffers. Similarly, for communication with off-chip memory, we choose to use FIFOs to facilitate communication between fused tasks, thereby simplifying the overlap. 
\newtext{Listing~\ref{lst:mm_inter} illustrates an example where the write function transfers data produced by Tasks 0 and 1 into a FIFO, which is then consumed by Task 5. As shown, the on-chip buffers used by Tasks 0 and 1 differ from those used by Task 5, allowing greater flexibility in choosing tile sizes for each task independently.}



\begin{figure}[!htb]
\begin{lstlisting}[label={lst:mm_inter},caption={{\textit{FIFO-Based Inter-Task Communication for Intermediate Matrix \texttt{E}}: The first function writes data from the local buffer \texttt{E} into a FIFO, while the second reads from the FIFO to populate the next computation stage's on-chip buffer.} },escapeinside={(*@}{@*)}]
void write_E_FT0(float E[10][32],
                 hls::stream<float16> &fifo_E_from_task1_to_task5, int j0,
                 int i0) {
#pragma HLS inline off
  if (j0 < 0 || i0 < 0) {
    return;
  }
  for (int d0 = 0; d0 < 10; d0++) {
    for (int d1 = 0; d1 < 32; d1 += 16) {
#pragma HLS pipeline II = 1
      float16 tmp_fifo;
      for (j = 0; j < 16; j++){
#pragma HLS unroll
          tmp_fifo[j] = E[d0][d1 + j];
      }
      fifo_E_from_task1_to_task5.write(tmp_fifo);
    }
  }
}

void read_E_FT2(float E[180][192],
                hls::stream<float16> &fifo_E_from_task1_to_task5, int i0,
                int j0) {
#pragma HLS inline off
  if (j0 > 0 || i0 >= 18) {
    return;
  }
  for (int d1_0 = 0; d1_0 < 6; d1_0++) {
    for (int d0 = 0; d0 < 10; d0++) {
      for (int d1_1 = 0; d1_1 < 32; d1_1 += 16) {
        int d1 = d1_0 * 32 + d1_1;
        float16 tmp_fifo = fifo_E_from_task1_to_task5.read();
        for (j = 0; j < 16; j++){
#pragma HLS unroll
            if (d1 + j < 192)
                E[d0 + i0 * 10][d1 + 0 + j] = tmp_fifo[j];
        }
      }
    }
  }
}

\end{lstlisting}
\end{figure}

\subsection{Intra-Task}  
Each intra-task, corresponding to an intra-tile selected by the NLP, 
is implemented as a fully unrolled, independent function without 
communication with off-chip memory. Data for these tasks 
resides entirely on-chip. 

If the task involves reduction loops, 
inter-tile reduction loops are integrated into the 
function and pipelined. While the pipeline initiation interval (II) 
is greater than 1 due to reduction dependencies, 
other operations in the statement are pipelined efficiently.

Padding is handled at the intra-tile level. 
Non-reduction loops remain unchanged, allowing computation 
of padding values without excessive resource usage. 
For reduction loops, full tiles are computed first, 
and the intra-tile loop is adjusted to handle padding for partial tiles accurately.

\subsection{Inter-Tile Loop}
For each fused task and each inter-tile loop, we generate independent functions to facilitate efficient double or triple buffering. Using information from the NLP, we determine which arrays are defined and which are transferred at each level of granularity. 
Once we identify the data being transferred, we implement double buffering if only reads or writes are involved, or triple buffering if both reads and writes are required. This strategy allows us to overlap the operations of reading, storing, and computing at the innermost level, optimizing both communication and computation overlap.

\subsection{Concurrent Execution}  
Using the \textit{dataflow} pragma, independent tasks execute concurrently without requiring manual rewrites. Computation in a receiving task begins as soon as its shared buffer contains sufficient data.

\subsection{SLR Management}  
The NLP determines the SLR ID for each task, specifying where it will be executed. \framework generates a separate C++ file for each SLR, and data transfers between SLRs are managed via \textit{ap\_axiu} streams, ensuring efficient communication and minimizing transfer overhead.

\subsection{Design Regeneration}

In cases where bitstream generation fails for a given design, we support automatic regeneration of the design. Several strategies can be applied, such as tightening resource constraints or reducing the maximum unrolling factor. By reducing the available resources, the design becomes smaller, which can help alleviate congestion. Thanks to the flexibility of our NLP-based approach, we can retain parts of the previous solution—such as the SLR assignment—and selectively restrict resources only for the specific task or group of tasks responsible for the congestion. This is the method we employ to regenerate designs when bitstream generation fails.

\section{Evaluation}
\label{sec:evaluation}

\subsection{Setup}
\label{sec:eval:setup}

We evaluated our method using kernels from Polybench/C 4.2.1 \cite{polybench-web} with medium-sized datasets and single-precision floating-point computations. The selected kernels represent both memory-intensive and computation-intensive scenarios. Medium problem sizes were chosen to balance demonstration of efficacy and the feasibility of time-consuming RTL evaluations. 
Due to Allo’s lack of automatic code generation, we limited our experiments to the subset of PolyBench kernels provided in its artifact evaluation package~\cite{allo_artifact}, as results for other kernels were not available. \newtext{We include a comparison with Sisyphus using the \textit{n}-madd kernels, where $n$ denotes the number of matrix additions. In the case of 2-madd, the result of the first addition is used as the input to the second. For 3-madd, the results of the first two additions are both used as inputs to the final addition.}


Table~\ref{table:polybenchdesc} lists the benchmark kernels used in our evaluation. For each kernel, we report the computational complexity
(in terms of the number of operations) and the memory complexity (in terms of footprint of the input/output data). Additionally, we
include the data reuse order, which reflects the approximate number of times one data element is reused across different
computations within the kernel. Kernels with reuse on the order of $\mathcal{O}({N})$, where $N$ denotes the problem size (e.g., the
number of rows or columns in an $N \times N$ input matrix), are typically considered compute-bound as their theoretical arithmetic intensity (assuming perfect reuse on-chip) is $\mathcal{O}({N})$. In contrast, kernels with
reuse on the order of $\mathcal{O}({1})$ are generally classified as memory-bound. In practice, compute-bound kernels require very careful bufferization on-chip to minimize off-chip communications and achieve the theoretical arithmetic intensity.
The final column (\textbf{Communication Between Tasks}) 
represents the number of data elements that are transferred between tasks in the dataflow design,
excluding any initial input loading or setup overhead.

\begin{table}[!htb]
\begin{center}
\footnotesize{
\begin{tabular}{l|l p{1.4cm} p{1.4cm} l >{\raggedleft\arraybackslash}p{2cm}}
\textsf{Benchmark} & \textsf{Description} & Ops \;\;\;\; Complexity & Mem \;\;\;\;  Complexity & Reuse & Comm. Between Tasks  \\ \hline
\texttt{bicg} &	BiCG sub-kernel of BiCGStab solver & $\mathcal{O}(N^2)$& $\mathcal{O}(N^2)$ &	$\mathcal{O}(1)$& 0 \\
\texttt{madd} & Matrix add. (C = A + B) & $\mathcal{O}(N^2)$& $\mathcal{O}(N^2)$ & $\mathcal{O}(1)$& 0 \\
\texttt{mvt} &	Matrix Vector product and Transpose & $\mathcal{O}(N^2) $ & $\mathcal{O}(N^2) $ &	$\mathcal{O}(1)$& 0\\
\midrule
\texttt{atax} &	Matrix transpose and vector mult. & $\mathcal{O}(N^2)$& $\mathcal{O}(N^2) $ & 	$\mathcal{O}(1)$& $N$ \\
\texttt{gesummv} &		Scalar, vector and matrix mult. & $\mathcal{O}(N^2)$& $\mathcal{O}(N^2)$ & $\mathcal{O}(1)$& $2N$ \\
\midrule
\texttt{2-madd} & 2 Matrix add. (D = (A + B) + C) & $\mathcal{O}(N^2) $ & $\mathcal{O}(N^2) $ & $\mathcal{O}(1)$&  $N^2$  \\
\texttt{3-madd} &  3 Matrix add. (F = (A + B) + (C + D)) & $\mathcal{O}(N^2) $ & $\mathcal{O}(N^2) $ & $\mathcal{O}(1)$& 2$N^2$ \\
\texttt{gemver} &	Vector mult. and matrix add. & $\mathcal{O}(N^2) $ & $\mathcal{O}(N^2)$ & $\mathcal{O}(1)$	& $2N^2+2N$ \\
\midrule

\texttt{2mm} &	2 Matrix Mult. ($\alpha A B C+\beta D$) & $\mathcal{O}(N^3)$& $\mathcal{O}(N^2) $ &	$\mathcal{O}(N)$& $N^2$  \\

\texttt{gemm} &	Matrix-multiply ($C=\alpha A B+\beta C$) & $\mathcal{O}(N^3) $ & $\mathcal{O}(N^2) $ &	$\mathcal{O}(N)$& $N^2$ \\

\texttt{syr2k} &		Symmetric rank-2k update & $\mathcal{O}(N^3) $ & $\mathcal{O}(N^2)$ & $\mathcal{O}(N)$ & $N^2$\\
\texttt{syrk} &	Symmetric rank-k update & $\mathcal{O}(N^3)$& $\mathcal{O}(N^2)$ & 	$\mathcal{O}(N)$& $N^2$\\

\texttt{trmm} &	Triangular matrix-mult. & $\mathcal{O}(N^3) $ & $\mathcal{O}(N^2) $ &	$\mathcal{O}(N)$& $N^2$ \\ 
\midrule
\texttt{3mm} &	3 Matrix Mult. ($(A B) (C D)$) & $\mathcal{O}(N^3) $ & $\mathcal{O}(N^2) $ & 	$\mathcal{O}(N)$& 2$N^2$  \\
\texttt{symm} &	Symmetric matrix-mult. & $\mathcal{O}(N^3)$ & $\mathcal{O}(N^2) $ &	$\mathcal{O}(N)$& $2N^2$\\
\end{tabular}
}
\caption{\label{table:polybenchdesc}Benchmark Kernels Used for Evaluation: Computational Complexity, Memory Requirements, Data Reuse, and Inter-Task Communication Analysis}
\end{center}
\end{table}

NLP problems were solved using the AMPL description language and the Gurobi solver (version 11.0.0) with the \textit{qp:nonconvex=2} option for non-convex quadratic objectives and constraints. Evaluations included RTL simulation and on-board execution using the Alveo U55C FPGA, with a targeted frequency of 220 MHz. RTL simulation provided accurate latency estimates, contrasting with the overly optimistic Vitis HLS reports that assume perfect task overlapping with dataflow pragma.
The generated code from the frameworks are compiled using AMD/Xilinx Vitis HLS 2023.2 using the Vitis flow \cite{flow_amd}. This flow assumes that data initially resides off-chip and has a default latency of 64 cycles to bring onto on-chip memory. 
All frameworks utilize "unsafe math" optimizations, enabling commutative/associative reduction operations at the expense of precision.

\subsection{Experimental Evaluation}

The objective of this evaluation is to demonstrate the capability of our framework to generate code with high QoR, whether for memory-bound or computation-bound kernels. To achieve this, we compare our work with Allo, ScaleHLS, Sisyphus, AutoDSE and Stream-HLS. 

For ScaleHLS \cite{scalehls}, Allo \cite{allo}, and Stream-HLS \cite{stream_hls}, their kernels assume that data is already present in on-chip memory. To ensure a fair comparison, we modified their code to incorporate off-chip to on-chip data transfers. However, Allo’s 2mm and 3mm kernels already include this transfer mechanism, requiring no modifications.
%
We left Sisyphus and AutoDSE unchanged, as they already optimize bit width according to the problem size.
%
We use AutoDSE with the bottleneck method, setting a DSE timeout of 1,000 minutes and an HLS synthesis timeout of 180 minutes per task.
For Allo-generated designs, we utilize kernels from their artifact repository since Allo does not employ a DSE \cite{allo_artifact}. 
Sisyphus generated designs are generated using parameters consistent with the paper \cite{sisyphus}. We observe some differences in Sisyphus results, as our evaluation is conducted on a different board. The design variations between boards may lead to differences in performance.

%
For RTL simulation, we assume that the frameworks can utilize all the resources on the U55C FPGA with a constraint of partitioning any array of 1,024 due to AMD/Xilinx limitations.
For on-board FPGA evaluations, we consider two scenarios. The first scenario utilizes 60\% of one Super Logic Region (SLR), equivalent to 20\% of total board resources, for all frameworks. The second scenario is unique to our framework, leveraging all three SLRs, with 60\% utilization per SLR.
Most frameworks are not place-and-route aware, leading to congestion issues that prevent bitstream generation. Limiting frameworks to one SLR increases the likelihood of meeting timing requirements. AutoDSE, for instance, lacks dataflow support and applies pragmas within a single function, making multi-SLR bitstream generation impossible without manual intervention to split the function across SLRs.
\newtext{If congestion persists within an SLR, we manually adjust the NLP constraints based on the specific congestion issue and regenerate the HLS-C++ code, a process that typically takes only a few minutes. 
As our NLP model includes many constraints, both global and per SLR. If congestion occurs, we can address it by adjusting the relevant constraint. For example, if the issue is due to resource overutilization, we can retain the previous NLP solution and simply tighten the resource utilization constraint for the specific SLR. It is worth noting that different regeneration strategies are possible, depending on whether we prioritize achieving a high-quality result (QoR) or obtaining a solution quickly. For now, this decision is left to the user.
This step is currently performed manually to better understand the root causes of congestion, but it could be automated through analysis of the compiler-generated log files.}

It is worth noting that there is a difference in results between RTL simulation and on-board evaluation. This difference is expected, as RTL simulation involves fewer constraints, allowing for more aggressive unrolling and full utilization of available resources. In contrast, on-board evaluation is more complex, and excessive unrolling can easily lead to congestion. Therefore, we compare RTL simulation results with those of other frameworks to demonstrate our performance improvements. More importantly, we show that our approach generates code that can actually be implemented on board.

\subsection{Comparison}
%
Table \ref{tab:rtl} presents the RTL simulation results for \framework, Sisyphus \cite{sisyphus}, AutoDSE \cite{autodse}, ScaleHLS \cite{scalehls}, and Allo \cite{allo}. 
The last lines show the average and geometric mean performance improvement (PI) of \framework across evaluated kernels.

\begin{table}[!htb]
\centering
\footnotesize{
\begin{tabular}{@{}lrrrrrr@{}}
\toprule
Kernel & Ours	&Sisyphus &	ScaleHLS&	Allo &	AutoDSE & Stream-HLS\\
\midrule
2mm&308.38&195.09&37.13&46.58&0.41*&150.27 \\
3mm&368.36&178.97&43.04&60.40*&1.74*&174.75 \\
Atax&3.56&2.32&1.58&1.96&1.97*&1.71 \\
Bicg&15.41&2.32&1.70&14.17&0.99*&1.73 \\
Gemm&419.14&227.09&40.53&37.50&110.81*&203.48 \\
Gesummv&10.21&2.28&1.78&8.85&1.98*&1.72 \\
Mvt&14.65&5.54&7.39&8.77&7.80*&13.31 \\
Symm&212.20&200.30&0.06&16.72&14.68*&N/A \\
Syr2k&267.31&149.89&0.08&41.15&12.35*&N/A \\
Syrk&158.25&105.59&0.27&20.57&23.16*&N/A \\
Trmm&193.47&166.72&0.07&5.10&0.02*&N/A \\

\midrule
PI (Avg) & 1.00x&2.39x&927.20x&8.58x&973.14x& 3.46x \\
PI (gmean) &1.00x & 2.03x&48.03x&4.92x&25.82x&2.71x  \\
\bottomrule
\vspace{0.0005cm}
\end{tabular}%
}
\caption{Throughput Comparison (in GF/s) of PolyBench Kernels Across Frameworks Using RTL Simulation}
\label{tab:rtl}
\end{table}

Results marked with $*$ are from the Vitis HLS report, showing minimum latency as an optimistic estimate. 
The RTL simulation for \textit{3mm} with Allo was incomplete after two days.
The N/A values in the table for Scale-HLS correspond to kernels that have at least one loop with a non-constant trip count. Stream-HLS does not handle these cases, so we cannot make a comparison.

\framework consistently achieves superior QoR across all evaluated kernels compared to other frameworks. While Sisyphus \cite{sisyphus}, designed primarily for computation-bound kernels, demonstrates competitive results for these kernels, it shows a weakness for \textit{2mm} and \textit{3mm}. This disparity is due to Sisyphus lacking concurrent execution capabilities for independent matrix multiplications. The performance advantage of \framework stems from both concurrent task execution and efficient overlapping of computation and communication.

%
Although for \textit{bicg}, Allo and \framework do not use the same code transformation, the results are similar. Allo retains the original code structure by permuting the reduction loop outermost. The non-reduction loop is fully unrolled, and the reduction loop is pipelined. \framework partially unrolls both loops and pipelines the reduction loop.






\begin{table}[!htb]
\centering
\footnotesize{
\begin{tabular}{
@{\hspace{+3mm}}
r |
@{\hspace{+3mm}}
r
@{\hspace{+3mm}}
r
@{\hspace{+3mm}}
r
@{\hspace{+3mm}}
r
@{\hspace{+3mm}}
r |
@{\hspace{+3mm}}
@{\hspace{+3mm}}
r
@{\hspace{+3mm}}
r
@{\hspace{+3mm}}
r
@{\hspace{+3mm}}
r
@{\hspace{+3mm}}
r
@{\hspace{+3mm}}
@{}
}
\toprule
& \multicolumn{5}{c}{Sisyphus} & \multicolumn{5}{c}{Prometheus} \\ 
 Kernel   & GF/s & BRAM (\%) & DSP (\%) & FF (\%) &  LUT (\%)   & GF/s & BRAM (\%) & DSP (\%) & FF (\%) &  LUT (\%)    \\
\midrule
Madd&1.96&1&29&22&38&2.71&3&7&10&11 \\
2-madd&3.89&1&58&59&91&8.99&6&14&21&24 \\
3-madd&3.91&2&58&75&117&14.00&9&21&31&37 \\
\midrule
2mm&195.09&3&73&49&72&308.38&19&71&41&63 \\
3mm&178.97&16&98&86&122&368.36&20&83&52&85 \\
Gemm&227.09&6&51&42&72&419.14&18&53&33&45 \\
Gemver&17.93&24&116&56&68&22.51&31&53&34&54 \\
Mvt&5.54&23&2&6&10&14.65&6&22&16&51 \\
\bottomrule
\end{tabular}%
}
\caption{RTL Evaluation of Performance and Resource Utilization for Sisyphus and Prometheus}
\label{tab:eval_red}
\end{table}

\newtext{Table \ref{tab:eval_red} presents a comparison between Prometheus and Sisyphus on additional kernels. As shown, Prometheus generally achieves higher throughput and better overall resource utilization, except for BRAM usage, which is higher due to the use of double buffering. Notably, the 3-madd kernel shows a significant performance gain, as it enables concurrent execution of independent tasks. While similar benefits are also observed for 2mm and 3mm, the improvement is less pronounced.}
\newtextt{Despite achieving a 2.65x speedup on Mvt, we observe that \framework consumes more resources than Sisyphus. This is because Sisyphus attempts to unroll the loop with a large unroll factor and apply pipelining, but the compiler fails to achieve an initiation interval (II) of 1 and instead settles for II = 36. As a result, although they lose performance, the larger II significantly reduces the resource usage.}

\myparagraph{On Board Evaluation}
Table \ref{tab:eval} presents the results for the on-board evaluation. 
The column $T(ms)$ indicates the kernel execution time in milliseconds, GF/s represents the throughput in Giga Floating Operations per second, and the resource usage is detailed in the thousands for both LUTs (L) and FFs. URAM utilization is excluded as no kernels use it. Column F (MHz) shows the frequency achieved by each design, with a target frequency of 220 MHz for all designs. 

For \textit{Bicg} and \textit{Atax}, targeting 60\% of resources on one SLR caused congestion, requiring regeneration with a 55\% constraint. The \textit{3mm} bitstream from AutoDSE succeeded only with a 15\% constraint.

\begin{table}[!htb]
\centering
\footnotesize{
\begin{tabular}{@{}l
@{\hspace{+3mm}}
l
@{\hspace{+3mm}}
r
@{\hspace{+3mm}}
r
@{\hspace{+3mm}}
r
@{\hspace{+3mm}}
r
@{\hspace{+3mm}}
r
@{\hspace{+3mm}}
r
@{\hspace{+3mm}}
r@{}}
\toprule
& Kernel  & T (ms) & GF/s & DSP & BRAM & L(K) & FF(K) & F (MHz)  \\
\midrule
\multirow{5}{*}{\rotatebox[origin=c]{90}{1 SLR}} 
\multirow{5}{*}{\rotatebox[origin=c]{90}{  Sisyphus}}
&2mm    & 1.20  &30.57  &556  &510  &213  &276 & 220 \\
&3mm    & 1.52  &29.89  &984  &611  &230  &300  & 220 \\
&Atax   & 0.62  &1.03  &173  &450  &240  &250  & 220 \\
&Bicg   & 0.63  &1.02  &173  &451  &238  &265  & 217 \\
\midrule
\multirow{5}{*}{\rotatebox[origin=c]{90}{1 SLR}} 
\multirow{5}{*}{\rotatebox[origin=c]{90}{ AutoDSE}}
&2mm    & 92.25  &0.40  &963&353.5&287&292  & 205 \\
&3mm    & 110.34  &0.41  &1117&470&278&306  & 220 \\
&Atax   & 2.88  &0.22  &452&630.5&170&212  & 220 \\
&Bicg   & 1.13  &0.56  &196&867.5&168&217  & 214 \\
\midrule
\multirow{5}{*}{\rotatebox[origin=c]{90}{1 SLR}} 
\multirow{5}{*}{\rotatebox[origin=c]{90}{Ours}}
&2mm    & 0.56  & 65.13  &1941  &635.5  &371  &454  & 216  \\
&3mm    & 0.87  &51.95  & 1551  &635.5  &342  &423  & 220 \\
&Atax   & 0.24  &2.62  & 1081  &533.5  &234  &287  & 184 \\
&Bicg   & 0.15  &4.04  & 732  &311.5  &250  &302  & 220 \\
\midrule
\multirow{5}{*}{\rotatebox[origin=c]{90}{3 SLR}} 
\multirow{5}{*}{\rotatebox[origin=c]{90}{Ours}}
&2mm    & 0.29  & 125.54  &2752  &546  &428  &549  & 220  \\
&3mm    & 0.34  &134.07  &4379  &600  &684  &840  & 207  \\
&Atax   & 0.20  &3.10  & 1823  &634.5  &405  &539  & 137 \\
&Bicg   & 0.14  &4.34  & 1226  &241  &291  &380  & 177 \\

\bottomrule
\end{tabular}%
}
\caption{On-Board Evaluation of Performance and Resource Utilization Across Frameworks and SLR Configurations}
\label{tab:eval}
\end{table}


\framework achieves a 77.16× performance improvement over AutoDSE, the state-of-the-art (SOTA) model-free approach. This gain is accompanied by an average increase in resource utilization: 2.38x more DSPs, 1.08x more BRAM, 1.36x more LUTs, and 1.42x more FFs, reflecting the trade-offs made to achieve such high efficiency.
When compared to Sisyphus, a recent NLP-based approach, \framework demonstrates a notable 2.59× performance boost, leveraging an average of 3.88× more DSPs, 1.04× more BRAM, 1.31× more LUTs, and 1.33× more FFs, illustrating its ability to deliver substantial improvements while effectively utilizing additional resources.
%


Table \ref{tab:info} shows the parameters found by the NLP.
We use the same name of the iterator as  Polybench  4.2.1 \cite{polybench-web} code. $Si$ represent the statement in position $i$ in the code.
The second column gives the statement fused inside the same tasks, the third column sets the fused task order and loop order found by the NLP for the fused task and the last column supplies the data-tile size found by the NLP, if the array is present in a different fused task the fused-task
(defined in the second column) 
is precise.

%
%
Permutations are evident in the implementations of \textit{3mm}, \textit{Atax}, and \textit{Bicg}. For \textit{2mm} and \textit{3mm}, the NLP opted to fully transfer array B instead of overlapping computation and load, as the load operation had higher latency than computation.
In \textit{2mm}, the NLP retains the original loop order. Array \textit{tmp} is transferred from the first task to the second, with both tasks iterating over the first dimension using loop $i$. This enables a $4 \times 32$ data tile to be sent to the second task, which starts computation once a $4 \times 192$ tile of \textit{tmp}(FT1) is filled.
%
%

\begin{table}[!htb]
\centering
\footnotesize{
\begin{tabular}{@{}p{0.06\linewidth}p{0.18\linewidth}p{0.13\linewidth}p{0.50\linewidth}@{}}
\toprule
  & Fused Statement & Loop $\;$ Order & Data Tile Size  \\
\midrule
2mm &
FT0: S0, S1, FT1: S2, S3&   FT0: i,j,k, FT1: i,j,k   
&  tmp(FT0): $4\times32$, B: $212\times192$, A: $4\times212$, D: $4\times32$, C:$192\times 32$, tmp(FT1): $4\times192$         
\\
3mm &
FT0: S0, S1, FT1: S2, S3, FT2: S4, S5
& FT0: i,j,k, FT1: j,i,k FT2: j,i,k 
&  
E(FT0): $9\times16$, 
A: $9\times 200$,

B: $200 \times 192$,
F(FT1): $10\times10$, 

C: $10\times 224$,
D: $224 \times 10$,
G: $9\times10$,
E(FT2): $9\times192$, 
F(FT2): $192\times10$, 
\\
Atax &
FT0: S1, S2, FT1: S0, S3
& FT0: i,j, FT1: j,i
& 
tmp(FT0): $56$, 
y: $16$,
A(FT0): $392\times416$,  
A(FT1): $392\times 16$, 
tmp(FT1): $392$
\\
Bicg &
FT0: S1, S2, FT1: S0, S3
&   FT0: j,i FT1: i,j
&   s: $16$, A (FT0): $416\times 16$, r: $416$, q: $10$, A (FT1): $10 \times 400$, p: $400$, 
\\

\bottomrule
\vspace{0.0005cm}
\end{tabular}%
}
\caption{Fusion, loop order and data-tile size found by the NLP for the kernel evaluated on Table \ref{tab:eval} for 1 SLR}
\label{tab:info}
\vspace{-0.6cm}
\end{table}

As other frameworks cannot generate bitstreams for 3 SLRs without human intervention, we compare results for 1 and 3 SLRs using our framework. For \textit{2mm} and \textit{3mm}, performance improves due to increased resource utilization. However, for \textit{atax} and \textit{bicg}, where the bottleneck is memory transfer between off-chip and on-chip rather than parallelism, the improvement is negligible.

\subsection{Scalability}

Our NLP solver includes the option to set a timeout, returning the best design found so far without guaranteeing optimality. This feature enables efficient exploration of the solution space while ensuring adherence to time constraints when necessary.


When analyzing the time required to solve the Non-Linear Programming (NLP) problem in Sisyphus, we observe that our framework, Prometheus, achieves similar solution times for most benchmarks. However, there is a notable exception: the 3mm benchmark, which times out after 4 hours in Sisyphus, while Prometheus successfully finds a solution.

This efficiency stems from Prometheus’ ability to explore a larger optimization space while still maintaining fast solution times. The key reason is that all optimization techniques are seamlessly integrated within the design space. This structured integration ensures that when a decision is made for one optimization parameter, it naturally constrains the possible choices for others, thereby reducing the search complexity.

In the case of 3mm, which has six statements (including initialization), Sisyphus evaluates the product of all possible permutations since it relies on shared buffers. In contrast, Prometheus employs dataflow, which requires preserving the order of data between tasks that communicate through FIFOs. As a result, although each statement could in principle lead to 2! or 3! loop permutations, many of these permutations become invalid under dataflow constraints when tasks must communicate via FIFOs. This substantially reduces the effective search space.
Consequently, even with a broader search space, Prometheus converges efficiently to a solution in just 21.37 seconds.

A detailed comparison of solution times can be found in Table \ref{tab:sisyphus_prometheus}, which highlights the impact of our approach in managing the design space effectively.

\begin{table}[!htb]
    \centering
    \renewcommand{\arraystretch}{1.2}
    \footnotesize{
    \begin{tabular}{lcc}
        \hline
        \textbf{Benchmark} & \textbf{Sisyphus} & \textbf{Prometheus} \\
        \hline
        2mm     & 22.37     & 15.98 \\
        3mm     & 14400.08  & 21.37 \\
        Atax    & 3.18      & 7.03 \\
        Bicg    & 3.17      & 5.09 \\
        Gemm    & 2.19      & 4.31 \\
        Gesummv & 1.57      & 8.07 \\
        Mvt     & 1.71      & 1.26 \\
        Symm    & 7.13      & 11.46 \\
        Syr2k   & 5.29      & 6.07 \\
        Syrk    & 2.56      & 3.94 \\
        Trmm    & 4.34      & 5.96 \\
        \hline
        \textbf{Average}  & 1313.96  & 8.23 \\
        \textbf{Geo Mean} & 7.95     & 6.50 \\
        \hline
    \end{tabular}
    }
    \caption{Time Required (in seconds) for the NLP Solver to Find an Optimal Solution: Sisyphus vs. Prometheus Across Benchmarks (Timeout Set at 14,400s)}
    \label{tab:sisyphus_prometheus}
\end{table}



\section{Related Work}
\label{sec:related_work}

Recent years have seen the development of numerous frameworks and tools addressing the challenges of optimizing FPGA accelerators. While many tackle specific subproblems—such as pipelining, tiling, or dataflow—few propose holistic solutions integrating all stages of the transformation and scheduling pipeline. We categorize the most relevant related works according to key technical axes addressed by our framework: dataflow generation, shared-buffer optimization, code transformation, and tiling with padding. We also compare against general-purpose frameworks when applicable.

\myparagraph{Dataflow -- }
Dataflow principles have been extensively studied in models such as Kahn Process Networks \cite{Kahn74}, Dennis Dataflow \cite{10.5555/647323.721501}, synchronous dataflow languages \cite{5009446,benveniste1994data}, and for FPGA applications \cite{8425390, 10.1145/1556444.1556449, 6573210, 6567545, dace, 9609930, heteroflow, allo}. 
DaCe \cite{dace} introduces Stateful DataFlow multiGraphs to separate program definition from optimization, enabling transformations like tiling and double-buffering, though requiring user intervention. 
Stream-HLS \cite{stream_hls} automatically generates dataflow; however, it assumes that data are already on-chip, thereby overlooking communication with off-chip memory. Additionally, it offers very limited opportunities for parallelism.
Flower \cite{9609930} automates FPGA dataflow development but limits parallelism. Frameworks like \cite{allo, heteroflow}, built on HeteroCL \cite{heterocl}, optimize data placement and compute scheduling for heterogeneous systems, maximizing data reuse and bandwidth utilization. Systolic arrays \cite{autosa, polysa, 9099977, 10.1145/3400302.3415644, 10.1145/3570928} offer efficient computation for specific patterns but lack generalization. Application-specific frameworks \cite{soda, darkroom, 7827654, 9370315, 9221526, 10058509, 9774727, 7965030} demonstrate dataflow advantages but do not generalize across domains.
RapidStream-TAPA \cite{10.1145/3609335,10.1145/3593025} enhances the performance of dataflow designs and automates SLR placement. However, it requires an optimized kernel with a dataflow structure as input.

\myparagraph{Shared Buffer -- }
Shared buffer utilization through HLS has been extensively explored using methods such as NLP-based pragma insertion \cite{nlp_dse_poster, nlp_dse, sisyphus}, bottleneck DSE \cite{autodse}, and GNN-based latency and resource estimation \cite{harp, sohrabizadeh2022gnn, 10.1145/3489517.3530409, 10.1145/3489517.3530629, 9256462, 9803218, 10.1145/3489517.3530408}. However, these approaches lack effective integration of dataflow optimization techniques.

\myparagraph{Code Transformation -- }
Code transformation has been explored for CPUs \cite{pluto, pouchet.11.popl, kruse2020autotuning, baghdadi2019tiramisu}, GPUs \cite{ppcg}, and FPGAs \cite{pouchet.fpga.13, zhaopolsca, zhao2021phism, li2014throughput, liu2016loop, liu2017polyhedral, 7160061, choi.iccad.18}. Pluto \cite{pluto} minimizes communication and improves locality but can limit parallelism. FPGA-specific adaptations \cite{pouchet.fpga.13} leverage FIFOs for overlapping communication and computation but are restricted in parallelism. Recent works \cite{zhaopolsca, zhao2021phism} selectively use Pluto for latency minimization while avoiding non-HLS-friendly code. 
While \cite{li2014throughput, liu2016loop, liu2017polyhedral,7160061, choi.iccad.18} focus on optimizing pipelining techniques, they do not address parallelism or the coordination of computation and communication overlap, which are crucial for our objectives.
The \cite{scalehls, heterocl,pylog,allo,hida} compilers perform code transformations and pragma insertion, their modifications are primarily heuristic and based on loop properties. The paper \cite{sisyphus} described in Section \ref{sec:motivation} generates design with high QoR, 
but the absence of dataflow utilization hinders concurrent task execution. Additionally, their approach avoids padding, limiting the unroll factor to divisors of the loop's trip count and constraining tiling space.

\myparagraph{Tiling and Padding -- } 
Tiling is essential for balancing computation and communication. While prior works \cite{8056810, 10.1145/3445814.3446759} use cost models to minimize communication, our approach extends this to reduce overall latency. Techniques like NLP-based tiling \cite{10.1145/3445814.3446759} focus on CPUs, while Wedler \cite{wedler} optimizes DNNs on GPUs by fusing operators, enhancing data reuse, and employing padding to prevent bank conflicts. Padding is well-studied for reducing cache misses \cite{10.1145/2980983.2908123, 752655} and improving memory transfers \cite{6164955}, but its use for varying unroll factors on FPGAs remains underexplored.

\myparagraph{Discussion --} In summary, while existing works address specific optimization aspects such as dataflow generation, pragma insertion, or tiling, they remain fragmented solutions. Prometheus distinguishes itself by integrating these axes into a unified optimization flow. In particular, it combines dataflow construction, shared-buffer optimization, and code transformation with tiling and padding in a single framework. Furthermore, Prometheus automates the generation of HLS-compatible C++ and OpenCL code and introduces SLR-aware scheduling. This holistic approach, together with public availability on GitHub, provides a significant advance over prior FPGA optimization frameworks.

\section{Scopes of Application}
\newtextt{
\framework{} operates on affine C/C++ programs—a deliberate and powerful design choice. This class of programs captures a vast number of real-world, compute-intensive workloads found in domains such as dense linear algebra, signal processing, image filtering, and many AI kernels. These patterns are not only ubiquitous but also performance-critical in modern FPGA applications, and our evaluation covers typical patterns of computation and data reuse occurring in these computations.}

\newtextt{
Through MLIR~\cite{mlir} and emerging tools like Allo and Stream-HLS, our framework can seamlessly target programs written in higher-level languages such as Python. By translating these to affine MLIR representations and regenerating C/C++ code \cite{polygeist}, we enable a broad spectrum of users—from ML practitioners to scientific programmers—to benefit from hardware acceleration without sacrificing optimization quality.}

\newtextt{
Affine computations can be \emph{exactly} analyzed at compile-time, at the level of every operation and data element accessed \cite{feautrier1992some,feautrier1991dataflow}, leading to accurate analysis of
loop bounds, memory access patterns, data dependencies, and data reuse. This enables precise performance modeling, aggressive design space pruning, and robust application of advanced HLS optimizations such as tiling, unrolling, pipelining, and memory partitioning—tasks that are notoriously difficult in general-purpose compilers or non-affine settings.}

Extending to non-affine constructs is possible, but would require significant compromises. For instance, in the presence of indirect accesses like $A[B[i]]$, memory access patterns are no longer statically analyzable, preventing efficient tiling or partitioning. 
One must either fall back on high-latency off-chip access or resort to over-approximations, e.g., conservatively load entire
arrays on-chip—approaches that often degrade parallelism or resource utilization. Our methodology remains applicable,
but the design space becomes potentially much more limited due to superfluous constraints arising from over-approximations of the program description.

To achieve full automation and high-performance, our framework relies on the ability to distribute statements in loop nests into distinct loop nests, creating tasks for each generated loop nests afterwards. This property is required for maximal efficiency of our method, and fits typical dense linear algebra kernels for example. But some applications may not expose such property: for example,
the
time-dependent stencil computations such as Jacobi-2D involve temporal dependencies that prevent full loop distribution, making them less amenable to end-to-end optimization within our current framework. However, the computation performed at each time step can still be optimized effectively using our methodology.

\section{Conclusion}


In this work, we introduced \framework, a holistic optimization framework that unifies loop transformations, task concurrency, computation-communication overlap, and hardware-aware scheduling for FPGA accelerators. By formulating the optimization process as a Non-Linear Programming (NLP) problem, Prometheus enables a structured exploration of the vast design space while considering hardware constraints, memory bandwidth, and task parallelism. 

Our framework addresses key limitations in existing methodologies, which often optimize only isolated aspects of FPGA acceleration. Prometheus surpasses these approaches by integrating SLR-aware scheduling, dynamic memory management, and hybrid execution models that effectively balance dataflow streaming and shared buffering. This enables more efficient utilization of FPGA resources, leading to significant improvements in latency, throughput, and scalability.

Through extensive performance evaluations on computation-bound kernels, we demonstrated that Prometheus outperforms state-of-the-art frameworks.
Furthermore, Prometheus' automatic design space exploration significantly reduces the manual effort required in FPGA development. By automatically generating HLS-C++ code, OpenCL host code, and FPGA bitstreams, the framework streamlines the deployment process, making FPGA acceleration more accessible to a broader range of applications.

\noindent In summary, Prometheus provides an innovative and effective approach to FPGA optimization, delivering high-performance solutions while minimizing the complexity of hardware design.

\section*{Acknowledgments}
 This work was supported in part by the NSF award \#CCF-2211557, the CDSC industrial partners and
the AMD/HACC Program. 
The authors would like to thank Prof. Luciano Lavagno and Dr. DJ Wang for their helpful discussions and valuable contributions to this research.

\bibliographystyle{ACM-Reference-Format}
\bibliography{bibs/iccad23,bibs/sp,bibs/lnp,bibs/refs,bibs/gabriel,bibs/ierefs,bibs/ics15}


\begin{thebibliography}{84}


\ifx \showCODEN    \undefined \def \showCODEN     #1{\unskip}     \fi
\ifx \showISBNx    \undefined \def \showISBNx     #1{\unskip}     \fi
\ifx \showISBNxiii \undefined \def \showISBNxiii  #1{\unskip}     \fi
\ifx \showISSN     \undefined \def \showISSN      #1{\unskip}     \fi
\ifx \showLCCN     \undefined \def \showLCCN      #1{\unskip}     \fi
\ifx \shownote     \undefined \def \shownote      #1{#1}          \fi
\ifx \showarticletitle \undefined \def \showarticletitle #1{#1}   \fi
\ifx \showURL      \undefined \def \showURL       {\relax}        \fi
\providecommand\bibfield[2]{#2}
\providecommand\bibinfo[2]{#2}
\providecommand\natexlab[1]{#1}
\providecommand\showeprint[2][]{arXiv:#2}

\bibitem[all(2024)]%
        {allo_artifact}
 \bibinfo{year}{2024}\natexlab{}.
\newblock \bibinfo{title}{Allo Artifact: \url{https://github.com/cornell-zhang/allo-pldi24-artifact}}.
\newblock
\urldef\tempurl%
\url{https://github.com/cornell-zhang/allo-pldi24-artifact}
\showURL{%
\tempurl}


\bibitem[Abella-Gonz\'{a}lez et~al\mbox{.}(2021)]%
        {polybench-python}
\bibfield{author}{\bibinfo{person}{Miguel~\'{A}. Abella-Gonz\'{a}lez}, \bibinfo{person}{Pedro Carollo-Fern\'{a}ndez}, \bibinfo{person}{Louis-No\"{e}l Pouchet}, \bibinfo{person}{Fabrice Rastello}, {and} \bibinfo{person}{Gabriel Rodr\'{\i}guez}.} \bibinfo{year}{2021}\natexlab{}.
\newblock \showarticletitle{PolyBench/Python: benchmarking Python environments with polyhedral optimizations}. In \bibinfo{booktitle}{\emph{Proceedings of the 30th ACM SIGPLAN International Conference on Compiler Construction}} (Virtual, Republic of Korea) \emph{(\bibinfo{series}{CC 2021})}. \bibinfo{publisher}{Association for Computing Machinery}, \bibinfo{address}{New York, NY, USA}, \bibinfo{pages}{59–70}.
\newblock
\showISBNx{9781450383257}
\href{https://doi.org/10.1145/3446804.3446842}{doi:\nolinkurl{10.1145/3446804.3446842}}


\bibitem[Abid et~al\mbox{.}(2013)]%
        {6573210}
\bibfield{author}{\bibinfo{person}{Mariem Abid}, \bibinfo{person}{Khaled Jerbi}, \bibinfo{person}{Mickaël Raulet}, \bibinfo{person}{Olivier Déforges}, {and} \bibinfo{person}{Mohamed Abid}.} \bibinfo{year}{2013}\natexlab{}.
\newblock \showarticletitle{System level synthesis of dataflow programs: HEVC decoder case study}. In \bibinfo{booktitle}{\emph{Proceedings of the 2013 Electronic System Level Synthesis Conference (ESLsyn)}}. \bibinfo{pages}{1--6}.
\newblock


\bibitem[AMD/Xilinx(2024)]%
        {flow_amd}
\bibfield{author}{\bibinfo{person}{AMD/Xilinx}.} \bibinfo{year}{2024}\natexlab{}.
\newblock \bibinfo{booktitle}{\emph{AMD/Xilinx Vitis 2023.2 Documentation}}.
\newblock
\urldef\tempurl%
\url{https://docs.amd.com/r/en-US/ug1399-vitis-hls/Target-Flow-Overview}
\showURL{%
\tempurl}
\newblock
\shownote{Accessed: 2025-01-06}.


\bibitem[Amiri et~al\mbox{.}(2021)]%
        {9609930}
\bibfield{author}{\bibinfo{person}{Puya Amiri}, \bibinfo{person}{Arsène Pérard-Gayot}, \bibinfo{person}{Richard Membarth}, \bibinfo{person}{Philipp Slusallek}, \bibinfo{person}{Roland Leißa}, {and} \bibinfo{person}{Sebastian Hack}.} \bibinfo{year}{2021}\natexlab{}.
\newblock \showarticletitle{FLOWER: A comprehensive dataflow compiler for high-level synthesis}. In \bibinfo{booktitle}{\emph{2021 International Conference on Field-Programmable Technology (ICFPT)}}. \bibinfo{pages}{1--9}.
\newblock
\href{https://doi.org/10.1109/ICFPT52863.2021.9609930}{doi:\nolinkurl{10.1109/ICFPT52863.2021.9609930}}


\bibitem[Bacis et~al\mbox{.}(2017)]%
        {7965030}
\bibfield{author}{\bibinfo{person}{Marco Bacis}, \bibinfo{person}{Giuseppe Natale}, \bibinfo{person}{Emanuele Del~Sozzo}, {and} \bibinfo{person}{Marco~Domenico Santambrogio}.} \bibinfo{year}{2017}\natexlab{}.
\newblock \showarticletitle{A pipelined and scalable dataflow implementation of convolutional neural networks on FPGA}. In \bibinfo{booktitle}{\emph{2017 IEEE International Parallel and Distributed Processing Symposium Workshops (IPDPSW)}}. \bibinfo{pages}{90--97}.
\newblock
\href{https://doi.org/10.1109/IPDPSW.2017.44}{doi:\nolinkurl{10.1109/IPDPSW.2017.44}}


\bibitem[Baghdadi et~al\mbox{.}(2019)]%
        {baghdadi2019tiramisu}
\bibfield{author}{\bibinfo{person}{Riyadh Baghdadi}, \bibinfo{person}{Jessica Ray}, \bibinfo{person}{Malek~Ben Romdhane}, \bibinfo{person}{Emanuele Del~Sozzo}, \bibinfo{person}{Abdurrahman Akkas}, \bibinfo{person}{Yunming Zhang}, \bibinfo{person}{Patricia Suriana}, \bibinfo{person}{Shoaib Kamil}, {and} \bibinfo{person}{Saman Amarasinghe}.} \bibinfo{year}{2019}\natexlab{}.
\newblock \showarticletitle{Tiramisu: A polyhedral compiler for expressing fast and portable code}. In \bibinfo{booktitle}{\emph{2019 IEEE/ACM International Symposium on Code Generation and Optimization (CGO)}}. IEEE, \bibinfo{pages}{193--205}.
\newblock


\bibitem[Bai et~al\mbox{.}(2022)]%
        {10.1145/3489517.3530629}
\bibfield{author}{\bibinfo{person}{Yunsheng Bai}, \bibinfo{person}{Atefeh Sohrabizadeh}, \bibinfo{person}{Yizhou Sun}, {and} \bibinfo{person}{Jason Cong}.} \bibinfo{year}{2022}\natexlab{}.
\newblock \showarticletitle{Improving GNN-based accelerator design automation with meta learning}. In \bibinfo{booktitle}{\emph{Proceedings of the 59th ACM/IEEE Design Automation Conference}} (San Francisco, California) \emph{(\bibinfo{series}{DAC '22})}. \bibinfo{publisher}{Association for Computing Machinery}, \bibinfo{address}{New York, NY, USA}, \bibinfo{pages}{1347–1350}.
\newblock
\showISBNx{9781450391429}
\href{https://doi.org/10.1145/3489517.3530629}{doi:\nolinkurl{10.1145/3489517.3530629}}


\bibitem[Basalama and Cong(2025)]%
        {stream_hls}
\bibfield{author}{\bibinfo{person}{Suhail Basalama} {and} \bibinfo{person}{Jason Cong}.} \bibinfo{year}{2025}\natexlab{}.
\newblock \showarticletitle{Stream-HLS: Towards Automatic Dataflow Acceleration}. In \bibinfo{booktitle}{\emph{Proceedings of the 2025 ACM/SIGDA International Symposium on Field Programmable Gate Arrays}} (Monterey, CA, USA) \emph{(\bibinfo{series}{FPGA '25})}. \bibinfo{publisher}{Association for Computing Machinery}, \bibinfo{address}{New York, NY, USA}, \bibinfo{pages}{103–114}.
\newblock
\showISBNx{9798400713965}
\href{https://doi.org/10.1145/3706628.3708878}{doi:\nolinkurl{10.1145/3706628.3708878}}


\bibitem[Basalama et~al\mbox{.}(2023)]%
        {10.1145/3570928}
\bibfield{author}{\bibinfo{person}{Suhail Basalama}, \bibinfo{person}{Atefeh Sohrabizadeh}, \bibinfo{person}{Jie Wang}, \bibinfo{person}{Licheng Guo}, {and} \bibinfo{person}{Jason Cong}.} \bibinfo{year}{2023}\natexlab{}.
\newblock \showarticletitle{FlexCNN: An End-to-end Framework for Composing CNN Accelerators on FPGA}.
\newblock \bibinfo{journal}{\emph{ACM Trans. Reconfigurable Technol. Syst.}} \bibinfo{volume}{16}, \bibinfo{number}{2}, Article \bibinfo{articleno}{23} (\bibinfo{date}{mar} \bibinfo{year}{2023}), \bibinfo{numpages}{32}~pages.
\newblock
\showISSN{1936-7406}
\href{https://doi.org/10.1145/3570928}{doi:\nolinkurl{10.1145/3570928}}


\bibitem[Ben-Nun et~al\mbox{.}(2019)]%
        {dace}
\bibfield{author}{\bibinfo{person}{Tal Ben-Nun}, \bibinfo{person}{Johannes de Fine~Licht}, \bibinfo{person}{Alexandros~N. Ziogas}, \bibinfo{person}{Timo Schneider}, {and} \bibinfo{person}{Torsten Hoefler}.} \bibinfo{year}{2019}\natexlab{}.
\newblock \showarticletitle{Stateful dataflow multigraphs: a data-centric model for performance portability on heterogeneous architectures}. In \bibinfo{booktitle}{\emph{Proceedings of the International Conference for High Performance Computing, Networking, Storage and Analysis}} (Denver, Colorado) \emph{(\bibinfo{series}{SC '19})}. \bibinfo{publisher}{Association for Computing Machinery}, \bibinfo{address}{New York, NY, USA}, Article \bibinfo{articleno}{81}, \bibinfo{numpages}{14}~pages.
\newblock
\showISBNx{9781450362290}
\href{https://doi.org/10.1145/3295500.3356173}{doi:\nolinkurl{10.1145/3295500.3356173}}


\bibitem[Benveniste et~al\mbox{.}(1994)]%
        {benveniste1994data}
\bibfield{author}{\bibinfo{person}{Albert Benveniste}, \bibinfo{person}{Paul Caspi}, \bibinfo{person}{Paul Le~Guernic}, {and} \bibinfo{person}{Nicolas Halbwachs}.} \bibinfo{year}{1994}\natexlab{}.
\newblock \showarticletitle{Data-flow synchronous languages}. In \bibinfo{booktitle}{\emph{A Decade of Concurrency Reflections and Perspectives: REX School/Symposium Noordwijkerhout, The Netherlands June 1--4, 1993 Proceedings}}. Springer, \bibinfo{pages}{1--45}.
\newblock


\bibitem[Bhattacharyya et~al\mbox{.}(2009)]%
        {10.1145/1556444.1556449}
\bibfield{author}{\bibinfo{person}{Shuvra~S. Bhattacharyya}, \bibinfo{person}{Gordon Brebner}, \bibinfo{person}{J\"{o}rn~W. Janneck}, \bibinfo{person}{Johan Eker}, \bibinfo{person}{Carl von Platen}, \bibinfo{person}{Marco Mattavelli}, {and} \bibinfo{person}{Micka\"{e}l Raulet}.} \bibinfo{year}{2009}\natexlab{}.
\newblock \showarticletitle{OpenDF: a dataflow toolset for reconfigurable hardware and multicore systems}.
\newblock \bibinfo{journal}{\emph{SIGARCH Comput. Archit. News}} \bibinfo{volume}{36}, \bibinfo{number}{5} (\bibinfo{date}{jun} \bibinfo{year}{2009}), \bibinfo{pages}{29–35}.
\newblock
\showISSN{0163-5964}
\href{https://doi.org/10.1145/1556444.1556449}{doi:\nolinkurl{10.1145/1556444.1556449}}


\bibitem[Bondhugula et~al\mbox{.}(2008)]%
        {pluto}
\bibfield{author}{\bibinfo{person}{Uday Bondhugula}, \bibinfo{person}{Albert Hartono}, \bibinfo{person}{J. Ramanujam}, {and} \bibinfo{person}{P. Sadayappan}.} \bibinfo{year}{2008}\natexlab{}.
\newblock \showarticletitle{{A Practical Automatic Polyhedral Parallelizer and Locality Optimizer}}. In \bibinfo{booktitle}{\emph{Proceedings of the 29th ACM SIGPLAN Conference on Programming Language Design and Implementation}} (Tucson, AZ, USA) \emph{(\bibinfo{series}{PLDI '08})}. \bibinfo{publisher}{Association for Computing Machinery}, \bibinfo{address}{New York, NY, USA}, \bibinfo{pages}{101–113}.
\newblock
\showISBNx{9781595938602}
\href{https://doi.org/10.1145/1375581.1375595}{doi:\nolinkurl{10.1145/1375581.1375595}}


\bibitem[Chen et~al\mbox{.}(2024)]%
        {allo}
\bibfield{author}{\bibinfo{person}{Hongzheng Chen}, \bibinfo{person}{Niansong Zhang}, \bibinfo{person}{Shaojie Xiang}, \bibinfo{person}{Zhichen Zeng}, \bibinfo{person}{Mengjia Dai}, {and} \bibinfo{person}{Zhiru Zhang}.} \bibinfo{year}{2024}\natexlab{}.
\newblock \showarticletitle{Allo: A Programming Model for Composable Accelerator Design}.
\newblock \bibinfo{journal}{\emph{Proc. ACM Program. Lang.}} \bibinfo{volume}{8}, \bibinfo{number}{PLDI}, Article \bibinfo{articleno}{171} (\bibinfo{date}{jun} \bibinfo{year}{2024}).
\newblock
\href{https://doi.org/10.1145/3656401}{doi:\nolinkurl{10.1145/3656401}}


\bibitem[Chen et~al\mbox{.}(1993)]%
        {241594}
\bibfield{author}{\bibinfo{person}{W.Y. Chen}, \bibinfo{person}{P.P. Chang}, \bibinfo{person}{T.M. Conte}, {and} \bibinfo{person}{W.W. Hwu}.} \bibinfo{year}{1993}\natexlab{}.
\newblock \showarticletitle{The effect of code expanding optimizations on instruction cache design}.
\newblock \bibinfo{journal}{\emph{IEEE Trans. Comput.}} \bibinfo{volume}{42}, \bibinfo{number}{9} (\bibinfo{year}{1993}), \bibinfo{pages}{1045--1057}.
\newblock
\href{https://doi.org/10.1109/12.241594}{doi:\nolinkurl{10.1109/12.241594}}


\bibitem[Chi et~al\mbox{.}(2018)]%
        {soda}
\bibfield{author}{\bibinfo{person}{Yuze Chi}, \bibinfo{person}{Jason Cong}, \bibinfo{person}{Peng Wei}, {and} \bibinfo{person}{Peipei Zhou}.} \bibinfo{year}{2018}\natexlab{}.
\newblock \showarticletitle{{SODA: Stencil with Optimized Dataflow Architecture}}. In \bibinfo{booktitle}{\emph{2018 IEEE/ACM International Conference on Computer-Aided Design (ICCAD)}}. \bibinfo{pages}{1--8}.
\newblock
\href{https://doi.org/10.1145/3240765.3240850}{doi:\nolinkurl{10.1145/3240765.3240850}}


\bibitem[Choi and Cong(2018)]%
        {choi.iccad.18}
\bibfield{author}{\bibinfo{person}{Young-kyu Choi} {and} \bibinfo{person}{Jason Cong}.} \bibinfo{year}{2018}\natexlab{}.
\newblock \showarticletitle{{HLS-Based Optimization and Design Space Exploration for Applications with Variable Loop Bounds}}. In \bibinfo{booktitle}{\emph{2018 IEEE/ACM International Conference on Computer-Aided Design (ICCAD)}}. \bibinfo{pages}{1--8}.
\newblock
\href{https://doi.org/10.1145/3240765.3240815}{doi:\nolinkurl{10.1145/3240765.3240815}}


\bibitem[Cong et~al\mbox{.}(2018)]%
        {DBLP:journals/corr/abs-1807-01340}
\bibfield{author}{\bibinfo{person}{Jason Cong}, \bibinfo{person}{Zhenman Fang}, \bibinfo{person}{Yuchen Hao}, \bibinfo{person}{Peng Wei}, \bibinfo{person}{Cody~Hao Yu}, \bibinfo{person}{Chen Zhang}, {and} \bibinfo{person}{Peipei Zhou}.} \bibinfo{year}{2018}\natexlab{}.
\newblock \showarticletitle{Best-Effort {FPGA} Programming: {A} Few Steps Can Go a Long Way}.
\newblock \bibinfo{journal}{\emph{CoRR}}  \bibinfo{volume}{abs/1807.01340} (\bibinfo{year}{2018}).
\newblock
\showeprint[arXiv]{1807.01340}
\urldef\tempurl%
\url{http://arxiv.org/abs/1807.01340}
\showURL{%
\tempurl}


\bibitem[Cong and Wang(2018)]%
        {polysa}
\bibfield{author}{\bibinfo{person}{Jason Cong} {and} \bibinfo{person}{Jie Wang}.} \bibinfo{year}{2018}\natexlab{}.
\newblock \showarticletitle{PolySA: Polyhedral-Based Systolic Array Auto-Compilation}. In \bibinfo{booktitle}{\emph{2018 IEEE/ACM International Conference on Computer-Aided Design (ICCAD)}}. \bibinfo{pages}{1--8}.
\newblock
\href{https://doi.org/10.1145/3240765.3240838}{doi:\nolinkurl{10.1145/3240765.3240838}}


\bibitem[de~Fine~Licht et~al\mbox{.}(2021)]%
        {9370315}
\bibfield{author}{\bibinfo{person}{Johannes de Fine~Licht}, \bibinfo{person}{Andreas Kuster}, \bibinfo{person}{Tiziano De~Matteis}, \bibinfo{person}{Tal Ben-Nun}, \bibinfo{person}{Dominic Hofer}, {and} \bibinfo{person}{Torsten Hoefler}.} \bibinfo{year}{2021}\natexlab{}.
\newblock \showarticletitle{StencilFlow: Mapping Large Stencil Programs to Distributed Spatial Computing Systems}. In \bibinfo{booktitle}{\emph{2021 IEEE/ACM International Symposium on Code Generation and Optimization (CGO)}}. \bibinfo{pages}{315--326}.
\newblock
\href{https://doi.org/10.1109/CGO51591.2021.9370315}{doi:\nolinkurl{10.1109/CGO51591.2021.9370315}}


\bibitem[Dennis(1974)]%
        {10.5555/647323.721501}
\bibfield{author}{\bibinfo{person}{J.~B. Dennis}.} \bibinfo{year}{1974}\natexlab{}.
\newblock \showarticletitle{First version of a data flow procedure language}. In \bibinfo{booktitle}{\emph{Programming Symposium, Proceedings Colloque Sur La Programmation}}. \bibinfo{publisher}{Springer-Verlag}, \bibinfo{address}{Berlin, Heidelberg}, \bibinfo{pages}{362–376}.
\newblock
\showISBNx{3540068597}


\bibitem[Denzler et~al\mbox{.}(2023)]%
        {10058509}
\bibfield{author}{\bibinfo{person}{Alain Denzler}, \bibinfo{person}{Geraldo~F. Oliveira}, \bibinfo{person}{Nastaran Hajinazar}, \bibinfo{person}{Rahul Bera}, \bibinfo{person}{Gagandeep Singh}, \bibinfo{person}{Juan Gómez-Luna}, {and} \bibinfo{person}{Onur Mutlu}.} \bibinfo{year}{2023}\natexlab{}.
\newblock \showarticletitle{Casper: Accelerating Stencil Computations Using Near-Cache Processing}.
\newblock \bibinfo{journal}{\emph{IEEE Access}}  \bibinfo{volume}{11} (\bibinfo{year}{2023}), \bibinfo{pages}{22136--22154}.
\newblock
\href{https://doi.org/10.1109/ACCESS.2023.3252002}{doi:\nolinkurl{10.1109/ACCESS.2023.3252002}}


\bibitem[Feautrier(1991)]%
        {feautrier1991dataflow}
\bibfield{author}{\bibinfo{person}{Paul Feautrier}.} \bibinfo{year}{1991}\natexlab{}.
\newblock \showarticletitle{Dataflow analysis of array and scalar references}.
\newblock \bibinfo{journal}{\emph{International Journal of Parallel Programming}} \bibinfo{volume}{20}, \bibinfo{number}{1} (\bibinfo{year}{1991}), \bibinfo{pages}{23--53}.
\newblock


\bibitem[Feautrier(1992)]%
        {feautrier1992some}
\bibfield{author}{\bibinfo{person}{Paul Feautrier}.} \bibinfo{year}{1992}\natexlab{}.
\newblock \showarticletitle{Some efficient solutions to the affine scheduling problem. Part II. Multidimensional time}.
\newblock \bibinfo{journal}{\emph{International journal of parallel programming}} \bibinfo{volume}{21}, \bibinfo{number}{6} (\bibinfo{year}{1992}), \bibinfo{pages}{389--420}.
\newblock


\bibitem[Grigoraş et~al\mbox{.}(2013)]%
        {6567545}
\bibfield{author}{\bibinfo{person}{Paul Grigoraş}, \bibinfo{person}{Xinyu Niu}, \bibinfo{person}{Jose G.~F. Coutinho}, \bibinfo{person}{Wayne Luk}, \bibinfo{person}{Jacob Bower}, {and} \bibinfo{person}{Oliver Pell}.} \bibinfo{year}{2013}\natexlab{}.
\newblock \showarticletitle{Aspect driven compilation for dataflow designs}. In \bibinfo{booktitle}{\emph{2013 IEEE 24th International Conference on Application-Specific Systems, Architectures and Processors}}. \bibinfo{pages}{18--25}.
\newblock
\href{https://doi.org/10.1109/ASAP.2013.6567545}{doi:\nolinkurl{10.1109/ASAP.2013.6567545}}


\bibitem[Guo et~al\mbox{.}(2023a)]%
        {10.1145/3609335}
\bibfield{author}{\bibinfo{person}{Licheng Guo}, \bibinfo{person}{Yuze Chi}, \bibinfo{person}{Jason Lau}, \bibinfo{person}{Linghao Song}, \bibinfo{person}{Xingyu Tian}, \bibinfo{person}{Moazin Khatti}, \bibinfo{person}{Weikang Qiao}, \bibinfo{person}{Jie Wang}, \bibinfo{person}{Ecenur Ustun}, \bibinfo{person}{Zhenman Fang}, \bibinfo{person}{Zhiru Zhang}, {and} \bibinfo{person}{Jason Cong}.} \bibinfo{year}{2023}\natexlab{a}.
\newblock \showarticletitle{TAPA: A Scalable Task-parallel Dataflow Programming Framework for Modern FPGAs with Co-optimization of HLS and Physical Design}.
\newblock \bibinfo{journal}{\emph{ACM Trans. Reconfigurable Technol. Syst.}} \bibinfo{volume}{16}, \bibinfo{number}{4}, Article \bibinfo{articleno}{63} (\bibinfo{date}{Dec.} \bibinfo{year}{2023}), \bibinfo{numpages}{31}~pages.
\newblock
\showISSN{1936-7406}
\href{https://doi.org/10.1145/3609335}{doi:\nolinkurl{10.1145/3609335}}


\bibitem[Guo et~al\mbox{.}(2023b)]%
        {10.1145/3593025}
\bibfield{author}{\bibinfo{person}{Licheng Guo}, \bibinfo{person}{Pongstorn Maidee}, \bibinfo{person}{Yun Zhou}, \bibinfo{person}{Chris Lavin}, \bibinfo{person}{Eddie Hung}, \bibinfo{person}{Wuxi Li}, \bibinfo{person}{Jason Lau}, \bibinfo{person}{Weikang Qiao}, \bibinfo{person}{Yuze Chi}, \bibinfo{person}{Linghao Song}, \bibinfo{person}{Yuanlong Xiao}, \bibinfo{person}{Alireza Kaviani}, \bibinfo{person}{Zhiru Zhang}, {and} \bibinfo{person}{Jason Cong}.} \bibinfo{year}{2023}\natexlab{b}.
\newblock \showarticletitle{RapidStream 2.0: Automated Parallel Implementation of Latency–Insensitive FPGA Designs Through Partial Reconfiguration}.
\newblock \bibinfo{journal}{\emph{ACM Trans. Reconfigurable Technol. Syst.}} \bibinfo{volume}{16}, \bibinfo{number}{4}, Article \bibinfo{articleno}{59} (\bibinfo{date}{Sept.} \bibinfo{year}{2023}), \bibinfo{numpages}{30}~pages.
\newblock
\showISSN{1936-7406}
\href{https://doi.org/10.1145/3593025}{doi:\nolinkurl{10.1145/3593025}}


\bibitem[Hegarty et~al\mbox{.}(2014)]%
        {darkroom}
\bibfield{author}{\bibinfo{person}{James Hegarty}, \bibinfo{person}{John Brunhaver}, \bibinfo{person}{Zachary DeVito}, \bibinfo{person}{Jonathan Ragan-Kelley}, \bibinfo{person}{Noy Cohen}, \bibinfo{person}{Steven Bell}, \bibinfo{person}{Artem Vasilyev}, \bibinfo{person}{Mark Horowitz}, {and} \bibinfo{person}{Pat Hanrahan}.} \bibinfo{year}{2014}\natexlab{}.
\newblock \showarticletitle{Darkroom: compiling high-level image processing code into hardware pipelines}.
\newblock \bibinfo{journal}{\emph{ACM Trans. Graph.}} \bibinfo{volume}{33}, \bibinfo{number}{4}, Article \bibinfo{articleno}{144} (\bibinfo{date}{jul} \bibinfo{year}{2014}), \bibinfo{numpages}{11}~pages.
\newblock
\showISSN{0730-0301}
\href{https://doi.org/10.1145/2601097.2601174}{doi:\nolinkurl{10.1145/2601097.2601174}}


\bibitem[Hong et~al\mbox{.}(2016)]%
        {10.1145/2980983.2908123}
\bibfield{author}{\bibinfo{person}{Changwan Hong}, \bibinfo{person}{Wenlei Bao}, \bibinfo{person}{Albert Cohen}, \bibinfo{person}{Sriram Krishnamoorthy}, \bibinfo{person}{Louis-No\"{e}l Pouchet}, \bibinfo{person}{Fabrice Rastello}, \bibinfo{person}{J. Ramanujam}, {and} \bibinfo{person}{P. Sadayappan}.} \bibinfo{year}{2016}\natexlab{}.
\newblock \showarticletitle{Effective padding of multidimensional arrays to avoid cache conflict misses}.
\newblock \bibinfo{journal}{\emph{SIGPLAN Not.}} \bibinfo{volume}{51}, \bibinfo{number}{6} (\bibinfo{date}{jun} \bibinfo{year}{2016}), \bibinfo{pages}{129–144}.
\newblock
\showISSN{0362-1340}
\href{https://doi.org/10.1145/2980983.2908123}{doi:\nolinkurl{10.1145/2980983.2908123}}


\bibitem[Huang et~al\mbox{.}(2021)]%
        {pylog}
\bibfield{author}{\bibinfo{person}{Sitao Huang}, \bibinfo{person}{Kun Wu}, \bibinfo{person}{Hyunmin Jeong}, \bibinfo{person}{Chengyue Wang}, \bibinfo{person}{Deming Chen}, {and} \bibinfo{person}{Wen-Mei Hwu}.} \bibinfo{year}{2021}\natexlab{}.
\newblock \showarticletitle{{PyLog: An Algorithm-Centric Python-Based FPGA Programming and Synthesis Flow}}.
\newblock \bibinfo{journal}{\emph{IEEE Trans. Comput.}} \bibinfo{volume}{70}, \bibinfo{number}{12} (\bibinfo{year}{2021}), \bibinfo{pages}{2015--2028}.
\newblock
\href{https://doi.org/10.1109/TC.2021.3123465}{doi:\nolinkurl{10.1109/TC.2021.3123465}}


\bibitem[Intel(2024)]%
        {intel_fpga}
\bibfield{author}{\bibinfo{person}{Intel}.} \bibinfo{year}{2024}\natexlab{}.
\newblock \bibinfo{title}{Intel}.
\newblock
\urldef\tempurl%
\url{https://www.intel.com/content/www/us/en/software/programmable/quartus-prime/hls-compiler.html}
\showURL{%
\tempurl}


\bibitem[Jia et~al\mbox{.}(2020)]%
        {9099977}
\bibfield{author}{\bibinfo{person}{Liancheng Jia}, \bibinfo{person}{Liqiang Lu}, \bibinfo{person}{Xuechao Wei}, {and} \bibinfo{person}{Yun Liang}.} \bibinfo{year}{2020}\natexlab{}.
\newblock \showarticletitle{Generating Systolic Array Accelerators With Reusable Blocks}.
\newblock \bibinfo{journal}{\emph{IEEE Micro}} \bibinfo{volume}{40}, \bibinfo{number}{4} (\bibinfo{year}{2020}), \bibinfo{pages}{85--92}.
\newblock
\href{https://doi.org/10.1109/MM.2020.2997611}{doi:\nolinkurl{10.1109/MM.2020.2997611}}


\bibitem[Kahn(1974)]%
        {Kahn74}
\bibfield{author}{\bibinfo{person}{Gilles Kahn}.} \bibinfo{year}{1974}\natexlab{}.
\newblock \showarticletitle{The Semantics of a Simple Language for Parallel Programming}. In \bibinfo{booktitle}{\emph{Information Processing, Proceedings of the 6th {IFIP} Congress 1974, Stockholm, Sweden, August 5-10, 1974}}, \bibfield{editor}{\bibinfo{person}{Jack~L. Rosenfeld}} (Ed.). \bibinfo{publisher}{North-Holland}, \bibinfo{pages}{471--475}.
\newblock


\bibitem[Khatti et~al\mbox{.}(2023)]%
        {pasta}
\bibfield{author}{\bibinfo{person}{Moazin Khatti}, \bibinfo{person}{Xingyu Tian}, \bibinfo{person}{Yuze Chi}, \bibinfo{person}{Licheng Guo}, \bibinfo{person}{Jason Cong}, {and} \bibinfo{person}{Zhenman Fang}.} \bibinfo{year}{2023}\natexlab{}.
\newblock \showarticletitle{PASTA: Programming and Automation Support for Scalable Task-Parallel HLS Programs on Modern Multi-Die FPGAs}. In \bibinfo{booktitle}{\emph{2023 IEEE 31st Annual International Symposium on Field-Programmable Custom Computing Machines (FCCM)}}. \bibinfo{pages}{12--22}.
\newblock
\href{https://doi.org/10.1109/FCCM57271.2023.00011}{doi:\nolinkurl{10.1109/FCCM57271.2023.00011}}


\bibitem[Korol et~al\mbox{.}(2022)]%
        {9774727}
\bibfield{author}{\bibinfo{person}{Guilherme Korol}, \bibinfo{person}{Michael~Guilherme Jordan}, \bibinfo{person}{Mateus~Beck Rutzig}, {and} \bibinfo{person}{Antonio Carlos~Schneider Beck}.} \bibinfo{year}{2022}\natexlab{}.
\newblock \showarticletitle{AdaFlow: A Framework for Adaptive Dataflow CNN Acceleration on FPGAs}. In \bibinfo{booktitle}{\emph{2022 Design, Automation \& Test in Europe Conference \& Exhibition (DATE)}}. \bibinfo{pages}{244--249}.
\newblock
\href{https://doi.org/10.23919/DATE54114.2022.9774727}{doi:\nolinkurl{10.23919/DATE54114.2022.9774727}}


\bibitem[Kruse et~al\mbox{.}(2020)]%
        {kruse2020autotuning}
\bibfield{author}{\bibinfo{person}{Michael Kruse}, \bibinfo{person}{Hal Finkel}, {and} \bibinfo{person}{Xingfu Wu}.} \bibinfo{year}{2020}\natexlab{}.
\newblock \showarticletitle{{Autotuning search space for loop transformations}}. In \bibinfo{booktitle}{\emph{2020 IEEE/ACM 6th Workshop on the LLVM Compiler Infrastructure in HPC (LLVM-HPC) and Workshop on Hierarchical Parallelism for Exascale Computing (HiPar)}}. IEEE, \bibinfo{pages}{12--22}.
\newblock


\bibitem[Lai et~al\mbox{.}(2019)]%
        {heterocl}
\bibfield{author}{\bibinfo{person}{Yi-Hsiang Lai}, \bibinfo{person}{Yuze Chi}, \bibinfo{person}{Yuwei Hu}, \bibinfo{person}{Jie Wang}, \bibinfo{person}{Cody~Hao Yu}, \bibinfo{person}{Yuan Zhou}, \bibinfo{person}{Jason Cong}, {and} \bibinfo{person}{Zhiru Zhang}.} \bibinfo{year}{2019}\natexlab{}.
\newblock \showarticletitle{{HeteroCL: A Multi-Paradigm Programming Infrastructure for Software-Defined Reconfigurable Computing}}. In \bibinfo{booktitle}{\emph{Proceedings of the 2019 ACM/SIGDA International Symposium on Field-Programmable Gate Arrays}} (Seaside, CA, USA) \emph{(\bibinfo{series}{FPGA '19})}. \bibinfo{publisher}{Association for Computing Machinery}, \bibinfo{address}{New York, NY, USA}, \bibinfo{pages}{242–251}.
\newblock
\showISBNx{9781450361378}
\href{https://doi.org/10.1145/3289602.3293910}{doi:\nolinkurl{10.1145/3289602.3293910}}


\bibitem[Lai et~al\mbox{.}(2020)]%
        {10.1145/3400302.3415644}
\bibfield{author}{\bibinfo{person}{Yi-Hsiang Lai}, \bibinfo{person}{Hongbo Rong}, \bibinfo{person}{Size Zheng}, \bibinfo{person}{Weihao Zhang}, \bibinfo{person}{Xiuping Cui}, \bibinfo{person}{Yunshan Jia}, \bibinfo{person}{Jie Wang}, \bibinfo{person}{Brendan Sullivan}, \bibinfo{person}{Zhiru Zhang}, \bibinfo{person}{Yun Liang}, \bibinfo{person}{Youhui Zhang}, \bibinfo{person}{Jason Cong}, \bibinfo{person}{Nithin George}, \bibinfo{person}{Jose Alvarez}, \bibinfo{person}{Christopher Hughes}, {and} \bibinfo{person}{Pradeep Dubey}.} \bibinfo{year}{2020}\natexlab{}.
\newblock \showarticletitle{SuSy: a programming model for productive construction of high-performance systolic arrays on FPGAs}. In \bibinfo{booktitle}{\emph{Proceedings of the 39th International Conference on Computer-Aided Design}} (Virtual Event, USA) \emph{(\bibinfo{series}{ICCAD '20})}. \bibinfo{publisher}{Association for Computing Machinery}, \bibinfo{address}{New York, NY, USA}, Article \bibinfo{articleno}{73}, \bibinfo{numpages}{9}~pages.
\newblock
\showISBNx{9781450380263}
\href{https://doi.org/10.1145/3400302.3415644}{doi:\nolinkurl{10.1145/3400302.3415644}}


\bibitem[Lattner et~al\mbox{.}(2021)]%
        {mlir}
\bibfield{author}{\bibinfo{person}{Chris Lattner}, \bibinfo{person}{Mehdi Amini}, \bibinfo{person}{Uday Bondhugula}, \bibinfo{person}{Albert Cohen}, \bibinfo{person}{Andy Davis}, \bibinfo{person}{Jacques Pienaar}, \bibinfo{person}{River Riddle}, \bibinfo{person}{Tatiana Shpeisman}, \bibinfo{person}{Nicolas Vasilache}, {and} \bibinfo{person}{Oleksandr Zinenko}.} \bibinfo{year}{2021}\natexlab{}.
\newblock \showarticletitle{{{MLIR}}: Scaling Compiler Infrastructure for Domain Specific Computation}. In \bibinfo{booktitle}{\emph{2021 {{IEEE/ACM}} International Symposium on Code Generation and Optimization (CGO)}}. \bibinfo{pages}{2--14}.
\newblock
\href{https://doi.org/10.1109/CGO51591.2021.9370308}{doi:\nolinkurl{10.1109/CGO51591.2021.9370308}}


\bibitem[Lee and Messerschmitt(1987)]%
        {5009446}
\bibfield{author}{\bibinfo{person}{Edward~Ashford Lee} {and} \bibinfo{person}{David~G. Messerschmitt}.} \bibinfo{year}{1987}\natexlab{}.
\newblock \showarticletitle{Static Scheduling of Synchronous Data Flow Programs for Digital Signal Processing}.
\newblock \bibinfo{journal}{\emph{IEEE Trans. Comput.}} \bibinfo{volume}{C-36}, \bibinfo{number}{1} (\bibinfo{year}{1987}), \bibinfo{pages}{24--35}.
\newblock
\href{https://doi.org/10.1109/TC.1987.5009446}{doi:\nolinkurl{10.1109/TC.1987.5009446}}


\bibitem[Li et~al\mbox{.}(2020)]%
        {heterohalide}
\bibfield{author}{\bibinfo{person}{Jiajie Li}, \bibinfo{person}{Yuze Chi}, {and} \bibinfo{person}{Jason Cong}.} \bibinfo{year}{2020}\natexlab{}.
\newblock \showarticletitle{{HeteroHalide: From Image Processing DSL to Efficient FPGA Acceleration}}. In \bibinfo{booktitle}{\emph{Proceedings of the 2020 ACM/SIGDA International Symposium on Field-Programmable Gate Arrays}} (Seaside, CA, USA) \emph{(\bibinfo{series}{FPGA '20})}. \bibinfo{publisher}{Association for Computing Machinery}, \bibinfo{address}{New York, NY, USA}, \bibinfo{pages}{51–57}.
\newblock
\showISBNx{9781450370998}
\href{https://doi.org/10.1145/3373087.3375320}{doi:\nolinkurl{10.1145/3373087.3375320}}


\bibitem[Li et~al\mbox{.}(2014)]%
        {li2014throughput}
\bibfield{author}{\bibinfo{person}{Peng Li}, \bibinfo{person}{Louis-No{\"e}l Pouchet}, {and} \bibinfo{person}{Jason Cong}.} \bibinfo{year}{2014}\natexlab{}.
\newblock \showarticletitle{{Throughput optimization for high-level synthesis using resource constraints}}. In \bibinfo{booktitle}{\emph{Int. Workshop on Polyhedral Compilation Techniques (IMPACT’14)}}.
\newblock


\bibitem[Li et~al\mbox{.}(2021)]%
        {10.1145/3445814.3446759}
\bibfield{author}{\bibinfo{person}{Rui Li}, \bibinfo{person}{Yufan Xu}, \bibinfo{person}{Aravind Sukumaran-Rajam}, \bibinfo{person}{Atanas Rountev}, {and} \bibinfo{person}{P. Sadayappan}.} \bibinfo{year}{2021}\natexlab{}.
\newblock \showarticletitle{Analytical characterization and design space exploration for optimization of CNNs}. In \bibinfo{booktitle}{\emph{Proceedings of the 26th ACM International Conference on Architectural Support for Programming Languages and Operating Systems}} (Virtual, USA) \emph{(\bibinfo{series}{ASPLOS '21})}. \bibinfo{publisher}{Association for Computing Machinery}, \bibinfo{address}{New York, NY, USA}, \bibinfo{pages}{928–942}.
\newblock
\showISBNx{9781450383172}
\href{https://doi.org/10.1145/3445814.3446759}{doi:\nolinkurl{10.1145/3445814.3446759}}


\bibitem[Liu et~al\mbox{.}(2015)]%
        {7160061}
\bibfield{author}{\bibinfo{person}{Junyi Liu}, \bibinfo{person}{Samuel Bayliss}, {and} \bibinfo{person}{George~A. Constantinides}.} \bibinfo{year}{2015}\natexlab{}.
\newblock \showarticletitle{{Offline Synthesis of Online Dependence Testing: Parametric Loop Pipelining for HLS}}. In \bibinfo{booktitle}{\emph{2015 IEEE 23rd Annual International Symposium on Field-Programmable Custom Computing Machines}}. \bibinfo{pages}{159--162}.
\newblock
\href{https://doi.org/10.1109/FCCM.2015.31}{doi:\nolinkurl{10.1109/FCCM.2015.31}}


\bibitem[Liu et~al\mbox{.}(2017b)]%
        {liu2017polyhedral}
\bibfield{author}{\bibinfo{person}{Junyi Liu}, \bibinfo{person}{John Wickerson}, \bibinfo{person}{Samuel Bayliss}, {and} \bibinfo{person}{George~A Constantinides}.} \bibinfo{year}{2017}\natexlab{b}.
\newblock \showarticletitle{{Polyhedral-based dynamic loop pipelining for high-level synthesis}}.
\newblock \bibinfo{journal}{\emph{IEEE Transactions on Computer-Aided Design of Integrated Circuits and Systems}} \bibinfo{volume}{37}, \bibinfo{number}{9} (\bibinfo{year}{2017}), \bibinfo{pages}{1802--1815}.
\newblock


\bibitem[Liu et~al\mbox{.}(2016)]%
        {liu2016loop}
\bibfield{author}{\bibinfo{person}{Junyi Liu}, \bibinfo{person}{John Wickerson}, {and} \bibinfo{person}{George~A Constantinides}.} \bibinfo{year}{2016}\natexlab{}.
\newblock \showarticletitle{{Loop splitting for efficient pipelining in high-level synthesis}}. In \bibinfo{booktitle}{\emph{2016 IEEE 24th Annual International Symposium on Field-Programmable Custom Computing Machines (FCCM)}}. IEEE, \bibinfo{pages}{72--79}.
\newblock


\bibitem[Liu et~al\mbox{.}(2017a)]%
        {8056810}
\bibfield{author}{\bibinfo{person}{Junyi Liu}, \bibinfo{person}{John Wickerson}, {and} \bibinfo{person}{George~A. Constantinides}.} \bibinfo{year}{2017}\natexlab{a}.
\newblock \showarticletitle{{Tile size selection for optimized memory reuse in high-level synthesis}}. In \bibinfo{booktitle}{\emph{2017 27th International Conference on Field Programmable Logic and Applications (FPL)}}. \bibinfo{pages}{1--8}.
\newblock
\href{https://doi.org/10.23919/FPL.2017.8056810}{doi:\nolinkurl{10.23919/FPL.2017.8056810}}


\bibitem[Microchip(2023)]%
        {legup}
\bibfield{author}{\bibinfo{person}{Microchip}.} \bibinfo{year}{2023}\natexlab{}.
\newblock \bibinfo{title}{{SmartHLS Compiler Software}}.
\newblock
\urldef\tempurl%
\url{https://www.microchip.com/en-us/products/fpgas-and-plds/fpga-and-soc-design-tools/smarthls-compiler}
\showURL{%
\tempurl}


\bibitem[Moses et~al\mbox{.}(2021)]%
        {polygeist}
\bibfield{author}{\bibinfo{person}{William~S. Moses}, \bibinfo{person}{Lorenzo Chelini}, \bibinfo{person}{Ruizhe Zhao}, {and} \bibinfo{person}{Oleksandr Zinenko}.} \bibinfo{year}{2021}\natexlab{}.
\newblock \showarticletitle{Polygeist: Raising C to Polyhedral MLIR}. In \bibinfo{booktitle}{\emph{Proceedings of the ACM International Conference on Parallel Architectures and Compilation Techniques}} (Virtual Event) \emph{(\bibinfo{series}{PACT '21})}. \bibinfo{publisher}{Association for Computing Machinery}, \bibinfo{address}{New York, NY, USA}, \bibinfo{numpages}{12}~pages.
\newblock


\bibitem[Natale et~al\mbox{.}(2016)]%
        {7827654}
\bibfield{author}{\bibinfo{person}{Giuseppe Natale}, \bibinfo{person}{Giulio Stramondo}, \bibinfo{person}{Pietro Bressana}, \bibinfo{person}{Riccardo Cattaneo}, \bibinfo{person}{Donatella Sciuto}, {and} \bibinfo{person}{Marco~D. Santambrogio}.} \bibinfo{year}{2016}\natexlab{}.
\newblock \showarticletitle{A polyhedral model-based framework for dataflow implementation on FPGA devices of Iterative Stencil Loops}. In \bibinfo{booktitle}{\emph{2016 IEEE/ACM International Conference on Computer-Aided Design (ICCAD)}}. \bibinfo{pages}{1--8}.
\newblock
\href{https://doi.org/10.1145/2966986.2966995}{doi:\nolinkurl{10.1145/2966986.2966995}}


\bibitem[Panda et~al\mbox{.}(1999)]%
        {752655}
\bibfield{author}{\bibinfo{person}{P.R. Panda}, \bibinfo{person}{H. Nakamura}, \bibinfo{person}{N.D. Dutt}, {and} \bibinfo{person}{A. Nicolau}.} \bibinfo{year}{1999}\natexlab{}.
\newblock \showarticletitle{Augmenting loop tiling with data alignment for improved cache performance}.
\newblock \bibinfo{journal}{\emph{IEEE Trans. Comput.}} \bibinfo{volume}{48}, \bibinfo{number}{2} (\bibinfo{year}{1999}), \bibinfo{pages}{142--149}.
\newblock
\href{https://doi.org/10.1109/12.752655}{doi:\nolinkurl{10.1109/12.752655}}


\bibitem[Peverelli et~al\mbox{.}(2018)]%
        {8425390}
\bibfield{author}{\bibinfo{person}{Francesco Peverelli}, \bibinfo{person}{Marco Rabozzi}, \bibinfo{person}{Emanuele Del~Sozzo}, {and} \bibinfo{person}{Marco~D. Santambrogio}.} \bibinfo{year}{2018}\natexlab{}.
\newblock \showarticletitle{OXiGen: A Tool for Automatic Acceleration of C Functions Into Dataflow FPGA-Based Kernels}. In \bibinfo{booktitle}{\emph{2018 IEEE International Parallel and Distributed Processing Symposium Workshops (IPDPSW)}}. \bibinfo{pages}{91--98}.
\newblock
\href{https://doi.org/10.1109/IPDPSW.2018.00023}{doi:\nolinkurl{10.1109/IPDPSW.2018.00023}}


\bibitem[PoCC({[n.\,d.]})]%
        {pocc-web}
PoCC \bibinfo{year}{[n.\,d.]}\natexlab{}.
\newblock \bibinfo{booktitle}{\emph{{PoCC}, the {P}olyhedral {C}ompiler {C}ollection 1.3}}.
\newblock
\urldef\tempurl%
\url{http://pocc.sourceforge.net}
\showURL{%
\tempurl}


\bibitem[Polybench({[n.\,d.]})]%
        {polybench}
Polybench \bibinfo{year}{[n.\,d.]}\natexlab{}.
\newblock \bibinfo{title}{PolyBench/C: the Polyhedral Benchmark suite}.
\newblock
\newblock
\shownote{\url{http://tinyurl.com/m7ztgex}}.


\bibitem[Pouchet et~al\mbox{.}(2011)]%
        {pouchet.11.popl}
\bibfield{author}{\bibinfo{person}{Louis-No\"{e}l Pouchet}, \bibinfo{person}{Uday Bondhugula}, \bibinfo{person}{C{\'e}dric Bastoul}, \bibinfo{person}{Albert Cohen}, \bibinfo{person}{J. Ramanujam}, \bibinfo{person}{P. Sadayappan}, {and} \bibinfo{person}{Nicolas Vasilache}.} \bibinfo{year}{2011}\natexlab{}.
\newblock \showarticletitle{Loop Transformations: Convexity, Pruning and Optimization}. In \bibinfo{booktitle}{\emph{Proceedings of the 38th Annual ACM SIGPLAN-SIGACT Symposium on Principles of Programming Languages}} (Austin, Texas, USA) \emph{(\bibinfo{series}{POPL '11})}. \bibinfo{publisher}{ACM}, \bibinfo{address}{New York, NY, USA}, \bibinfo{pages}{549--562}.
\newblock
\showISBNx{978-1-4503-0490-0}
\href{https://doi.org/10.1145/1926385.1926449}{doi:\nolinkurl{10.1145/1926385.1926449}}


\bibitem[Pouchet and Yuki(2024)]%
        {polybench-web}
\bibfield{author}{\bibinfo{person}{Louis-Noel Pouchet} {and} \bibinfo{person}{Tomofumi Yuki}.} \bibinfo{year}{Retrieved 2024}\natexlab{}.
\newblock \bibinfo{title}{Polybench: The polyhedral benchmark suite}.
\newblock
\urldef\tempurl%
\url{http://polybench.sourceforge.net}
\showURL{%
\tempurl}


\bibitem[Pouchet et~al\mbox{.}(2013a)]%
        {pouchet:fpga13}
\bibfield{author}{\bibinfo{person}{Louis-No{\"e}l Pouchet}, \bibinfo{person}{Peng Zhang}, \bibinfo{person}{P. Sadayappan}, {and} \bibinfo{person}{Jason Cong}.} \bibinfo{year}{2013}\natexlab{a}.
\newblock \showarticletitle{Polyhedral-Based Data Reuse Optimization for Configurable Computing}. In \bibinfo{booktitle}{\emph{21st ACM/SIGDA International Symposium on Field-Programmable Gate Arrays (FPGA'13)}}. \bibinfo{publisher}{ACM Press}, \bibinfo{address}{Monterey, California}.
\newblock


\bibitem[Pouchet et~al\mbox{.}(2013b)]%
        {pouchet.fpga.13}
\bibfield{author}{\bibinfo{person}{Louis-Noel Pouchet}, \bibinfo{person}{Peng Zhang}, \bibinfo{person}{P. Sadayappan}, {and} \bibinfo{person}{Jason Cong}.} \bibinfo{year}{2013}\natexlab{b}.
\newblock \showarticletitle{{Polyhedral-Based Data Reuse Optimization for Configurable Computing}}. In \bibinfo{booktitle}{\emph{Proceedings of the ACM/SIGDA International Symposium on Field Programmable Gate Arrays}} (Monterey, California, USA) \emph{(\bibinfo{series}{FPGA '13})}. \bibinfo{publisher}{Association for Computing Machinery}, \bibinfo{address}{New York, NY, USA}, \bibinfo{pages}{29–38}.
\newblock
\showISBNx{9781450318877}
\href{https://doi.org/10.1145/2435264.2435273}{doi:\nolinkurl{10.1145/2435264.2435273}}


\bibitem[Pouget et~al\mbox{.}(2024)]%
        {nlp_dse_poster}
\bibfield{author}{\bibinfo{person}{St\'{e}phane Pouget}, \bibinfo{person}{Louis-No\"{e}l Pouchet}, {and} \bibinfo{person}{Jason Cong}.} \bibinfo{year}{2024}\natexlab{}.
\newblock \showarticletitle{Automatic Hardware Pragma Insertion in High-Level Synthesis: A Non-Linear Programming Approach}. In \bibinfo{booktitle}{\emph{Proceedings of the 2024 ACM/SIGDA International Symposium on Field Programmable Gate Arrays}} (Monterey,CA,USA) \emph{(\bibinfo{series}{FPGA '24})}. \bibinfo{publisher}{Association for Computing Machinery}, \bibinfo{address}{New York, NY, USA}, \bibinfo{pages}{184}.
\newblock
\showISBNx{9798400704185}
\href{https://doi.org/10.1145/3626202.3637593}{doi:\nolinkurl{10.1145/3626202.3637593}}


\bibitem[Pouget et~al\mbox{.}(2025a)]%
        {nlp_dse}
\bibfield{author}{\bibinfo{person}{St\'{e}phane Pouget}, \bibinfo{person}{Louis-No\"{e}l Pouchet}, {and} \bibinfo{person}{Jason Cong}.} \bibinfo{year}{2025}\natexlab{a}.
\newblock \showarticletitle{Automatic Hardware Pragma Insertion in High-Level Synthesis: A Non-Linear Programming Approach}.
\newblock \bibinfo{journal}{\emph{ACM Trans. Des. Autom. Electron. Syst.}} \bibinfo{volume}{30}, \bibinfo{number}{2}, Article \bibinfo{articleno}{26} (\bibinfo{date}{Feb.} \bibinfo{year}{2025}), \bibinfo{numpages}{44}~pages.
\newblock
\showISSN{1084-4309}
\href{https://doi.org/10.1145/3711847}{doi:\nolinkurl{10.1145/3711847}}


\bibitem[Pouget et~al\mbox{.}(2025b)]%
        {sisyphus}
\bibfield{author}{\bibinfo{person}{St\'{e}phane Pouget}, \bibinfo{person}{Louis-No\"{e}l Pouchet}, {and} \bibinfo{person}{Jason Cong}.} \bibinfo{year}{2025}\natexlab{b}.
\newblock \showarticletitle{A Unified Framework for Automated Code Transformation and Pragma Insertion}. In \bibinfo{booktitle}{\emph{Proceedings of the 2025 ACM/SIGDA International Symposium on Field Programmable Gate Arrays}} (Monterey, CA, USA) \emph{(\bibinfo{series}{FPGA '25})}. \bibinfo{publisher}{Association for Computing Machinery}, \bibinfo{address}{New York, NY, USA}, \bibinfo{pages}{187–198}.
\newblock
\showISBNx{9798400713965}
\href{https://doi.org/10.1145/3706628.3708873}{doi:\nolinkurl{10.1145/3706628.3708873}}


\bibitem[Shi et~al\mbox{.}(2023)]%
        {wedler}
\bibfield{author}{\bibinfo{person}{Yining Shi}, \bibinfo{person}{Zhi Yang}, \bibinfo{person}{Jilong Xue}, \bibinfo{person}{Lingxiao Ma}, \bibinfo{person}{Yuqing Xia}, \bibinfo{person}{Ziming Miao}, \bibinfo{person}{Yuxiao Guo}, \bibinfo{person}{Fan Yang}, {and} \bibinfo{person}{Lidong Zhou}.} \bibinfo{year}{2023}\natexlab{}.
\newblock \showarticletitle{Welder: Scheduling Deep Learning Memory Access via Tile-graph}. In \bibinfo{booktitle}{\emph{17th USENIX Symposium on Operating Systems Design and Implementation (OSDI 23)}}. \bibinfo{publisher}{USENIX Association}, \bibinfo{address}{Boston, MA}, \bibinfo{pages}{701--718}.
\newblock
\showISBNx{978-1-939133-34-2}
\urldef\tempurl%
\url{https://www.usenix.org/conference/osdi23/presentation/shi}
\showURL{%
\tempurl}


\bibitem[Siemens(2023)]%
        {catapult}
\bibfield{author}{\bibinfo{person}{Siemens}.} \bibinfo{year}{2023}\natexlab{}.
\newblock \bibinfo{title}{{Catapult High-Level Synthesis}}.
\newblock
\urldef\tempurl%
\url{https://eda.sw.siemens.com/en-US/ic/catapult-high-level-synthesis/}
\showURL{%
\tempurl}


\bibitem[Singh et~al\mbox{.}(2020)]%
        {9221526}
\bibfield{author}{\bibinfo{person}{Gagandeep Singh}, \bibinfo{person}{Dionysios Diamantopoulos}, \bibinfo{person}{Christoph Hagleitner}, \bibinfo{person}{Juan Gomez-Luna}, \bibinfo{person}{Sander Stuijk}, \bibinfo{person}{Onur Mutlu}, {and} \bibinfo{person}{Henk Corporaal}.} \bibinfo{year}{2020}\natexlab{}.
\newblock \showarticletitle{NERO: A Near High-Bandwidth Memory Stencil Accelerator for Weather Prediction Modeling}. In \bibinfo{booktitle}{\emph{2020 30th International Conference on Field-Programmable Logic and Applications (FPL)}}. \bibinfo{pages}{9--17}.
\newblock
\href{https://doi.org/10.1109/FPL50879.2020.00014}{doi:\nolinkurl{10.1109/FPL50879.2020.00014}}


\bibitem[Sohrabizadeh et~al\mbox{.}(2022a)]%
        {sohrabizadeh2022gnn}
\bibfield{author}{\bibinfo{person}{Atefeh Sohrabizadeh}, \bibinfo{person}{Yunsheng Bai}, \bibinfo{person}{Yizhou Sun}, {and} \bibinfo{person}{Jason Cong}.} \bibinfo{year}{2022}\natexlab{a}.
\newblock \showarticletitle{{Automated Accelerator Optimization Aided by Graph Neural Networks}}. In \bibinfo{booktitle}{\emph{2022 59th ACM/IEEE Design Automation Conference (DAC)}}.
\newblock


\bibitem[Sohrabizadeh et~al\mbox{.}(2022b)]%
        {10.1145/3489517.3530409}
\bibfield{author}{\bibinfo{person}{Atefeh Sohrabizadeh}, \bibinfo{person}{Yunsheng Bai}, \bibinfo{person}{Yizhou Sun}, {and} \bibinfo{person}{Jason Cong}.} \bibinfo{year}{2022}\natexlab{b}.
\newblock \showarticletitle{Automated accelerator optimization aided by graph neural networks}. In \bibinfo{booktitle}{\emph{Proceedings of the 59th ACM/IEEE Design Automation Conference}} (San Francisco, California) \emph{(\bibinfo{series}{DAC '22})}. \bibinfo{publisher}{Association for Computing Machinery}, \bibinfo{address}{New York, NY, USA}, \bibinfo{pages}{55–60}.
\newblock
\showISBNx{9781450391429}
\href{https://doi.org/10.1145/3489517.3530409}{doi:\nolinkurl{10.1145/3489517.3530409}}


\bibitem[Sohrabizadeh et~al\mbox{.}(2023)]%
        {harp}
\bibfield{author}{\bibinfo{person}{Atefeh Sohrabizadeh}, \bibinfo{person}{Yunsheng Bai}, \bibinfo{person}{Yizhou Sun}, {and} \bibinfo{person}{Jason Cong}.} \bibinfo{year}{2023}\natexlab{}.
\newblock \showarticletitle{Robust GNN-Based Representation Learning for HLS}. In \bibinfo{booktitle}{\emph{2023 IEEE/ACM International Conference on Computer Aided Design (ICCAD)}}. \bibinfo{pages}{1--9}.
\newblock
\href{https://doi.org/10.1109/ICCAD57390.2023.10323853}{doi:\nolinkurl{10.1109/ICCAD57390.2023.10323853}}


\bibitem[Sohrabizadeh et~al\mbox{.}(2021)]%
        {autodse}
\bibfield{author}{\bibinfo{person}{Atefeh Sohrabizadeh}, \bibinfo{person}{Cody~Hao Yu}, \bibinfo{person}{Min Gao}, {and} \bibinfo{person}{Jason Cong}.} \bibinfo{year}{2021}\natexlab{}.
\newblock \showarticletitle{{AutoDSE: Enabling Software Programmers Design Efficient FPGA Accelerators}}. In \bibinfo{booktitle}{\emph{The 2021 ACM/SIGDA International Symposium on Field-Programmable Gate Arrays}} (Virtual Event, USA) \emph{(\bibinfo{series}{FPGA '21})}. \bibinfo{publisher}{Association for Computing Machinery}, \bibinfo{address}{New York, NY, USA}, \bibinfo{pages}{147}.
\newblock
\showISBNx{9781450382182}
\href{https://doi.org/10.1145/3431920.3439464}{doi:\nolinkurl{10.1145/3431920.3439464}}


\bibitem[Ustun et~al\mbox{.}(2020)]%
        {9256462}
\bibfield{author}{\bibinfo{person}{Ecenur Ustun}, \bibinfo{person}{Chenhui Deng}, \bibinfo{person}{Debjit Pal}, \bibinfo{person}{Zhijing Li}, {and} \bibinfo{person}{Zhiru Zhang}.} \bibinfo{year}{2020}\natexlab{}.
\newblock \showarticletitle{Accurate Operation Delay Prediction for FPGA HLS Using Graph Neural Networks}. In \bibinfo{booktitle}{\emph{2020 IEEE/ACM International Conference On Computer Aided Design (ICCAD)}}. \bibinfo{pages}{1--9}.
\newblock


\bibitem[Verdoolaege(2011)]%
        {iscc}
\bibfield{author}{\bibinfo{person}{Sven Verdoolaege}.} \bibinfo{year}{2011}\natexlab{}.
\newblock \showarticletitle{Counting affine calculator and applications}. In \bibinfo{booktitle}{\emph{First International Workshop on Polyhedral Compilation Techniques (IMPACT’11), Chamonix, France}}.
\newblock


\bibitem[Verdoolaege et~al\mbox{.}(2013)]%
        {ppcg}
\bibfield{author}{\bibinfo{person}{Sven Verdoolaege}, \bibinfo{person}{Juan Carlos~Juega}, \bibinfo{person}{Albert Cohen}, \bibinfo{person}{Jos\'{e} Ignacio~G\'{o}mez}, \bibinfo{person}{Christian Tenllado}, {and} \bibinfo{person}{Francky Catthoor}.} \bibinfo{year}{2013}\natexlab{}.
\newblock \showarticletitle{Polyhedral parallel code generation for CUDA}.
\newblock \bibinfo{journal}{\emph{ACM Trans. Archit. Code Optim.}} \bibinfo{volume}{9}, \bibinfo{number}{4}, Article \bibinfo{articleno}{54} (\bibinfo{date}{jan} \bibinfo{year}{2013}), \bibinfo{numpages}{23}~pages.
\newblock
\showISSN{1544-3566}
\href{https://doi.org/10.1145/2400682.2400713}{doi:\nolinkurl{10.1145/2400682.2400713}}


\bibitem[Wang et~al\mbox{.}(2021)]%
        {autosa}
\bibfield{author}{\bibinfo{person}{Jie Wang}, \bibinfo{person}{Licheng Guo}, {and} \bibinfo{person}{Jason Cong}.} \bibinfo{year}{2021}\natexlab{}.
\newblock \showarticletitle{{AutoSA: A Polyhedral Compiler for High-Performance Systolic Arrays on FPGA}}. In \bibinfo{booktitle}{\emph{The 2021 ACM/SIGDA International Symposium on Field-Programmable Gate Arrays}} (Virtual Event, USA) \emph{(\bibinfo{series}{FPGA '21})}. \bibinfo{publisher}{Association for Computing Machinery}, \bibinfo{address}{New York, NY, USA}, \bibinfo{pages}{93–104}.
\newblock
\showISBNx{9781450382182}
\href{https://doi.org/10.1145/3431920.3439292}{doi:\nolinkurl{10.1145/3431920.3439292}}


\bibitem[Wang et~al\mbox{.}(2012)]%
        {6164955}
\bibfield{author}{\bibinfo{person}{Yuxin Wang}, \bibinfo{person}{Peng Zhang}, \bibinfo{person}{Xu Cheng}, {and} \bibinfo{person}{Jason Cong}.} \bibinfo{year}{2012}\natexlab{}.
\newblock \showarticletitle{An integrated and automated memory optimization flow for FPGA behavioral synthesis}. In \bibinfo{booktitle}{\emph{17th Asia and South Pacific Design Automation Conference}}. \bibinfo{pages}{257--262}.
\newblock
\href{https://doi.org/10.1109/ASPDAC.2012.6164955}{doi:\nolinkurl{10.1109/ASPDAC.2012.6164955}}


\bibitem[Wu et~al\mbox{.}(2022a)]%
        {9803218}
\bibfield{author}{\bibinfo{person}{Nan Wu}, \bibinfo{person}{Yuan Xie}, {and} \bibinfo{person}{Cong Hao}.} \bibinfo{year}{2022}\natexlab{a}.
\newblock \showarticletitle{{IronMan-Pro: Multi-objective Design Space Exploration in HLS via Reinforcement Learning and Graph Neural Network based Modeling}}.
\newblock \bibinfo{journal}{\emph{IEEE Transactions on Computer-Aided Design of Integrated Circuits and Systems}} (\bibinfo{year}{2022}), \bibinfo{pages}{1--1}.
\newblock
\href{https://doi.org/10.1109/TCAD.2022.3185540}{doi:\nolinkurl{10.1109/TCAD.2022.3185540}}


\bibitem[Wu et~al\mbox{.}(2023)]%
        {ironman}
\bibfield{author}{\bibinfo{person}{Nan Wu}, \bibinfo{person}{Yuan Xie}, {and} \bibinfo{person}{Cong Hao}.} \bibinfo{year}{2023}\natexlab{}.
\newblock \showarticletitle{{IronMan-Pro: Multiobjective Design Space Exploration in HLS via Reinforcement Learning and Graph Neural Network-Based Modeling}}.
\newblock \bibinfo{journal}{\emph{IEEE Transactions on Computer-Aided Design of Integrated Circuits and Systems}} \bibinfo{volume}{42}, \bibinfo{number}{3} (\bibinfo{year}{2023}), \bibinfo{pages}{900--913}.
\newblock
\href{https://doi.org/10.1109/TCAD.2022.3185540}{doi:\nolinkurl{10.1109/TCAD.2022.3185540}}


\bibitem[Wu et~al\mbox{.}(2022b)]%
        {10.1145/3489517.3530408}
\bibfield{author}{\bibinfo{person}{Nan Wu}, \bibinfo{person}{Hang Yang}, \bibinfo{person}{Yuan Xie}, \bibinfo{person}{Pan Li}, {and} \bibinfo{person}{Cong Hao}.} \bibinfo{year}{2022}\natexlab{b}.
\newblock \showarticletitle{High-level synthesis performance prediction using GNNs: benchmarking, modeling, and advancing}. In \bibinfo{booktitle}{\emph{Proceedings of the 59th ACM/IEEE Design Automation Conference}} (San Francisco, California) \emph{(\bibinfo{series}{DAC '22})}. \bibinfo{publisher}{Association for Computing Machinery}, \bibinfo{address}{New York, NY, USA}, \bibinfo{pages}{49–54}.
\newblock
\showISBNx{9781450391429}
\href{https://doi.org/10.1145/3489517.3530408}{doi:\nolinkurl{10.1145/3489517.3530408}}


\bibitem[Xiang et~al\mbox{.}(2022)]%
        {heteroflow}
\bibfield{author}{\bibinfo{person}{Shaojie Xiang}, \bibinfo{person}{Yi-Hsiang Lai}, \bibinfo{person}{Yuan Zhou}, \bibinfo{person}{Hongzheng Chen}, \bibinfo{person}{Niansong Zhang}, \bibinfo{person}{Debjit Pal}, {and} \bibinfo{person}{Zhiru Zhang}.} \bibinfo{year}{2022}\natexlab{}.
\newblock \showarticletitle{HeteroFlow: An Accelerator Programming Model with Decoupled Data Placement for Software-Defined FPGAs}. In \bibinfo{booktitle}{\emph{Proceedings of the 2022 ACM/SIGDA International Symposium on Field-Programmable Gate Arrays}} (Virtual Event, USA) \emph{(\bibinfo{series}{FPGA '22})}. \bibinfo{publisher}{Association for Computing Machinery}, \bibinfo{address}{New York, NY, USA}, \bibinfo{pages}{78–88}.
\newblock
\showISBNx{9781450391498}
\href{https://doi.org/10.1145/3490422.3502369}{doi:\nolinkurl{10.1145/3490422.3502369}}


\bibitem[Xilinx(0232)]%
        {vitis}
\bibfield{author}{\bibinfo{person}{AMD Xilinx}.} \bibinfo{year}{2023.2}\natexlab{}.
\newblock \bibinfo{title}{Vitis}.
\newblock
\urldef\tempurl%
\url{https://www.xilinx.com/products/design-tools/vitis.html}
\showURL{%
\tempurl}


\bibitem[Ye et~al\mbox{.}(2024)]%
        {hida}
\bibfield{author}{\bibinfo{person}{Hanchen Ye}, \bibinfo{person}{Hyegang Jun}, {and} \bibinfo{person}{Deming Chen}.} \bibinfo{year}{2024}\natexlab{}.
\newblock \showarticletitle{HIDA: A Hierarchical Dataflow Compiler for High-Level Synthesis}. In \bibinfo{booktitle}{\emph{Proceedings of the 29th ACM International Conference on Architectural Support for Programming Languages and Operating Systems, Volume 1}} (La Jolla,CA,USA) \emph{(\bibinfo{series}{ASPLOS '24})}. \bibinfo{publisher}{Association for Computing Machinery}, \bibinfo{address}{New York, NY, USA}, \bibinfo{pages}{215–230}.
\newblock
\showISBNx{9798400703720}
\href{https://doi.org/10.1145/3617232.3624850}{doi:\nolinkurl{10.1145/3617232.3624850}}


\bibitem[Ye et~al\mbox{.}(2022)]%
        {scalehls}
\bibfield{author}{\bibinfo{person}{Hanchen Ye}, \bibinfo{person}{HyeGang Jun}, \bibinfo{person}{Hyunmin Jeong}, \bibinfo{person}{Stephen Neuendorffer}, {and} \bibinfo{person}{Deming Chen}.} \bibinfo{year}{2022}\natexlab{}.
\newblock \showarticletitle{{ScaleHLS: A Scalable High-Level Synthesis Framework with Multi-Level Transformations and Optimizations: Invited}}. In \bibinfo{booktitle}{\emph{Proceedings of the 59th ACM/IEEE Design Automation Conference}} (San Francisco, California) \emph{(\bibinfo{series}{DAC '22})}. \bibinfo{publisher}{Association for Computing Machinery}, \bibinfo{address}{New York, NY, USA}, \bibinfo{pages}{1355–1358}.
\newblock
\showISBNx{9781450391429}
\href{https://doi.org/10.1145/3489517.3530631}{doi:\nolinkurl{10.1145/3489517.3530631}}


\bibitem[Zhang et~al\mbox{.}(2024)]%
        {pom}
\bibfield{author}{\bibinfo{person}{Weichuang Zhang}, \bibinfo{person}{Jieru Zhao}, \bibinfo{person}{Guan Shen}, \bibinfo{person}{Quan Chen}, \bibinfo{person}{Chen Chen}, {and} \bibinfo{person}{Minyi Guo}.} \bibinfo{year}{2024}\natexlab{}.
\newblock \showarticletitle{{ An Optimizing Framework on MLIR for Efficient FPGA-based Accelerator Generation }}. In \bibinfo{booktitle}{\emph{2024 IEEE International Symposium on High-Performance Computer Architecture (HPCA)}}. \bibinfo{publisher}{IEEE Computer Society}, \bibinfo{address}{Los Alamitos, CA, USA}, \bibinfo{pages}{75--90}.
\newblock
\href{https://doi.org/10.1109/HPCA57654.2024.00017}{doi:\nolinkurl{10.1109/HPCA57654.2024.00017}}


\bibitem[Zhao and Cheng(2021)]%
        {zhao2021phism}
\bibfield{author}{\bibinfo{person}{Ruizhe Zhao} {and} \bibinfo{person}{Jianyi Cheng}.} \bibinfo{year}{2021}\natexlab{}.
\newblock \showarticletitle{{Phism: Polyhedral High-Level Synthesis in MLIR}}.
\newblock \bibinfo{journal}{\emph{arXiv preprint arXiv:2103.15103}} (\bibinfo{year}{2021}).
\newblock


\bibitem[Zhao et~al\mbox{.}(2022)]%
        {zhaopolsca}
\bibfield{author}{\bibinfo{person}{Ruizhe Zhao}, \bibinfo{person}{Jianyi Cheng}, \bibinfo{person}{Wayne Luk}, {and} \bibinfo{person}{George~A Constantinides}.} \bibinfo{year}{2022}\natexlab{}.
\newblock \showarticletitle{{POLSCA: Polyhedral High-Level Synthesis with Compiler Transformations}}.
\newblock \bibinfo{journal}{\emph{arXiv}} (\bibinfo{year}{2022}).
\newblock


\end{thebibliography}

\end{document}